\documentclass[11pt]{article}

\usepackage{fullpage}
\usepackage{latexsym}
\usepackage{amsfonts}
\usepackage{amssymb}
\usepackage{amsmath}
\usepackage{color, graphicx}
\usepackage{hyperref}

\renewcommand{\bar}{\overline}
\renewcommand\Re{\operatorname{Re}}
\renewcommand\Im{\operatorname{Im}}

\newcommand{\beq}{\begin{equation}}
\newcommand{\eeq}{\end{equation}}
\newcommand{\ba}{\begin{array}}
\newcommand{\ea}{\end{array}}
\newcommand{\bea}{\begin{eqnarray*}}
\newcommand{\eea}{\end{eqnarray*}}
\newcommand{\bc}{\begin{center}}
\newcommand{\ec}{\end{center}}
\newcommand{\bt}{\begin{table}}
\newcommand{\et}{\end{table}}

\newcommand{\la}[1]{\label{#1}}

\newcommand{\no}{\noindent}

\newcommand{\rf}[1]{(\ref{#1})}
\newcommand{\beqno}{\begin{displaymath}}
\newcommand{\eeqno}{\end{displaymath}}

\newcommand{\been}{\begin{enumerate}}
\newcommand{\een}{\end{enumerate}}

\newcommand{\ra}{\rightarrow}

\newcommand{\ud}{\,\mathrm{d}}
\newcommand{\CC}{\mathbb{C}}

\newlength{\myheight}
\newlength{\mylength}

\newcounter{saveeqn}

\newtheorem{prop}{Proposition}

\def\XXint#1#2#3{{\setbox0=\hbox{$#1{#2#3}{\int}$}
     \vcenter{\hbox{$#2#3$}}\kern-.5\wd0}}

\allowdisplaybreaks[1]

\DeclareMathOperator{\erf}{erf}

\begin{document}

\title{Non-steady state heat conduction in composite walls}

\author{
Bernard Deconinck$^\dagger$~~Beatrice Pelloni$^*$~~Natalie Sheils$^\dagger$\\
~\\
$^\dagger$ Department of Applied Mathematics\\
University of Washington\\
Seattle, WA 98195-2420\\
bernard{@}amath.washington.edu~~nsheils{@}amath.washington.edu\\
~\\
$^*$ Department of Mathematics\\
University of Reading\\
RG6 6AX, UK\\
b.pelloni{@}reading.ac.uk}

\date{\emph{Accepted for publication\\ Proceedings of the Royal Society A}}

\maketitle

\begin{abstract}
The problem of heat conduction in one-dimensional piecewise homogeneous composite materials is examined by providing an explicit solution of the one-dimensional heat equation in each domain. The location of the interfaces is known, but neither temperature nor heat flux are prescribed there. Instead, the physical assumptions of their continuity at the interfaces are the only conditions imposed. The problem of two semi-infinite domains and that of two finite-sized domains are examined in detail. We indicate also how to extend the solution method to the setting of one finite-sized domain surrounded on both sides by semi-infinite domains, and on that of three finite-sized domains.
\end{abstract}

\section{Introduction}

The problem of heat conduction in a composite wall is a classical problem in design and construction. It is usual to restrict to the case of walls whose constitutive parts are in perfect thermal contact and have physical properties that are constant throughout the material and that are considered to be of infinite extent in the directions parallel to the wall. Further, we assume that temperature and heat flux do not vary in these directions. In that case, the mathematical model for heat conduction in each wall layer is given by \cite[Chapter~10]{HahnO}:

\begin{subequations}

\begin{align}\la{heatgeneral}
&&u_t^{(j)}&=\kappa_j u_{xx}^{(j)}, & x\in (a_j,b_j),\\
&&u^{(j)}(x,t=0)&=u_0^{(j)}(x), & x\in (a_j,b_j),
\end{align}

\end{subequations}

\no where $u^{(j)}(x,t)$ denotes the temperature in the wall layer indexed by $(j)$, $\kappa_j>0$ is the heat-conduction coefficient of the $j$-th layer (the inverse of its thermal diffusivity), $x=a_j$ is the left extent of the layer, and $x=b_j$ is its right extent. The sub-indices denote derivatives with respect to the one-dimensional spatial variable $x$ and the temporal variable $t$. The function $u^{(j)}_0(x)$ is the prescribed initial condition of the system. The continuity of the temperature $u^{(j)}$ and of its associated heat flux $\kappa_j u^{(j)}_x$ are imposed across the interface between layers. In what follows it is convenient to use the quantity $\sigma_j$, defined as the positive square root of $\kappa_j$: $\sigma_j=\sqrt{\kappa_j}$.

If the layer is either at the far left or far right of the wall, Dirichlet, Neumann, or Robin boundary conditions can be imposed on its far left or right boundary respectively, corresponding to prescribing ``outside'' temperature, heat flux, or a combination of these. A derivation of the interface boundary conditions is found in \cite[Chapter~1]{HahnO}. It should be noted that the set-up presented in \rf{heatgeneral} also applies to the case of one-dimensional rods in thermal contact.

In this paper, we use the Fokas Method \cite{DeconinckTrogdonVasan, FokasBook, FokasPelloni4} to provide explicit solution formulae for different heat transport interface problems of the type described above. We investigate problems in both finite and infinite domains and we compare our method with classical solution approaches that can be found in the literature. Throughout, our emphasis is on non-steady state solutions. Even for the simplest of the problems we consider (Section \ref{sec:2f}, two finite walls in thermal contact), the classical approach using separation of variables \cite{HahnO} can provide an explicit answer only implicitly.  Indeed, the solution obtained in \cite{HahnO} depends on certain eigenvalues defined through a transcendental equation that can be solved only numerically. In contrast, the Fokas Method produces an explicit solution formula involving only known quantities. For other problems we consider, to our knowledge no solution has been derived using classical methods, and we believe the solution formulae presented here are new.

The representation formulae for the solution can be evaluated numerically, hence the problem can be solved in practice using hybrid analytical-numerical approaches \cite{FlyerFokas} or asymptotic approximations for them may be obtained using standard techniques \cite{FokasBook}. The result of such a numerical calculation is shown at the end of Section~\ref{sec:ii}.

The problem of heat conduction through composite walls is discussed in many excellent texts, see for instance \cite{CarslawJaeger, HahnO}. References to the treatment of specific problems are given in the sections below where these problems are investigated. In Section~\ref{sec:ii} we investigate the problem of two semi-infinite walls. Section \ref{sec:2f} discusses the interface problem with two finite walls. Following that, we consider first the problem of one finite wall between two semi-infinite ones, and the problem of three finite walls. Both of these are briefly sketched in Section \ref{sec:other} and full solutions are presented in the electronic supplementary material.

\section{Two semi-infinite domains}\la{sec:ii}

In this section, we consider the problem of heat flow through two walls of semi-infinite width, or of two semi-infinite rods.

\smallskip
We seek two functions
$$
u^L(x,t),~\;x\in(-\infty,0),~ \;t\geq0,\qquad \qquad u^R(x,t),\;~~x\in(0,\infty),\;~~t\geq 0,
$$
satisfying the equations
\begin{subequations}\label{ueqns}
\begin{align}
u^L_t(x,t)&=\sigma_L^2u^L_{xx}(x,t),&x\in(-\infty,0),\;t>0,\\
u^R_t(x,t)&=\sigma_R^2u^R_{xx}(x,t),&x\in(0,\infty),\;t>0,
\end{align}
\end{subequations}
the initial conditions
\begin{subequations}\label{uic}
\begin{align}
u^L(x,0)&=u^L_0(x),&x\in(-\infty,0),\\
u^R(x,0)&=u^R_0(x),&x\in(0,\infty),
\end{align}
\end{subequations}
the asymptotic conditions
\begin{subequations}\la{ubcs}
\begin{align}
\lim_{x\to -\infty}u^L(x,t)&=\gamma_L, &t\geq0,\\
\lim_{x\to\infty}u^R(x,t)&=\gamma_R, &t\geq 0,
\end{align}
\end{subequations}
and the continuity interface conditions
\begin{subequations}\label{bcs}
\begin{align}
u^L(0,t)&=u^R(0,t), &t>0,\label{bcs1}\\
\sigma_L^2u^L_x(0,t)&=\sigma_R^2u^R_x(0,t), &t>0. \label{2i_bcs2}
\end{align}
\end{subequations}

\smallskip

The sub- and super-indices $L$ and $R$ denote the left and right rod, respectively. A special case of this problem is discussed in Chapter 10 of \cite{HahnO}, but only for a specific initial condition. Further, for the problem treated there both $\lim_{x\to\infty}u^R(x,t)$ and $\lim_{x\to -\infty}u^L(x,t)$ are assumed to be zero. This assumption is made for mathematical convenience and no physical reason exists to impose it. If constant (in time) limit values are assumed, a simple translation allows one of the limit values to be equated to zero, but not both. Since no great advantage is obtained by assuming a zero limit using our approach, we make the more general assumption~\eqref{ubcs}.

We define $v^L(x,t)=u^L(x,t)-\gamma^L$ and $v^R(x,t)=u^R(x,t)-\gamma^R$.  Then $v^L(x,t)$ and $v^R(x,t)$ satisfy

\begin{subequations}
\begin{align}
v^L_t(x,t)&=\sigma_L^2v^L_{xx}(x,t),&&x\in(-\infty,0)&&t\geq0,\label{heat1}\\
v^R_t(x,t)&=\sigma_R^2v^R_{xx}(x,t),&&x\in(0,\infty)&&t\geq0,\la{heat2} \\
\lim_{x\to-\infty}v^L(x,t)&=0,&&t\geq0,\\
\lim_{x\to\infty}v^R(x,t)&=0,&&t\geq 0,\\
v^L(0,t)+\gamma^L&=v^R(0,t)+\gamma^R,&&t\geq 0,\\
\sigma_L^2v^L_x(0,t)&=\sigma_R^2v^R_x(0,t),&&t\geq 0.
\end{align}
\end{subequations}

At this point, we start by following the standard steps in the application of the Fokas Method~\cite{DeconinckTrogdonVasan, FokasBook, FokasPelloni4}. We begin with the so-called ``local relations''~\cite{DeconinckTrogdonVasan}

\begin{subequations}
\begin{align}
\label{localLnz}
(e^{-ikx+(\sigma_Lk)^2 t}v^L(x,t))_t&=(\sigma_L^2e^{-ikx+(\sigma_Lk)^2 t}(v^L_x(x,t)+ik v^L(x,t)))_x,\\
\label{localRnz}
(e^{-ikx+(\sigma_R k)^2 t}v^R(x,t))_t&=(\sigma_R^2 e^{-ikx+(\sigma_R k)^2 t}(v^R_x(x,t)+ik v^R(x,t)))_x.
\end{align}
\end{subequations}
These relations are a one-parameter family obtained by rewriting \eqref{heat1} and \eqref{heat2}.

\begin{figure}[htbp]
\begin{center}
\def\svgwidth{4in}
   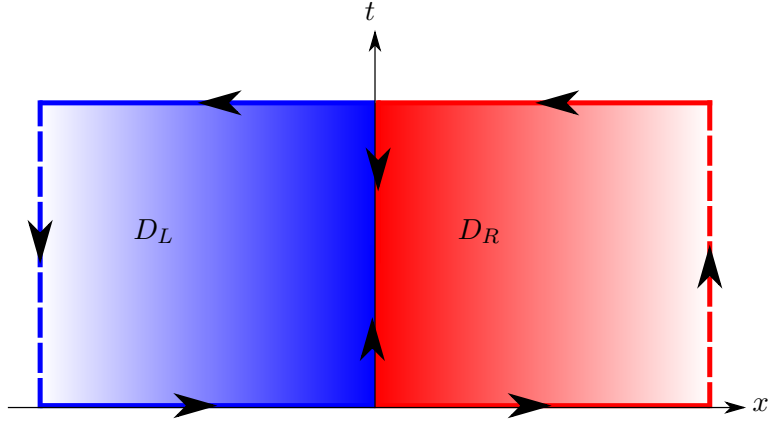 
   \caption{Domains for the application of Green's formula for $v^L(x,t)$ and $v^R(x,t)$.   \label{fig:LHP_RHP}}
  \end{center}
\end{figure}

Applying Green's formula~\cite{AblowitzFokas} in the strip $(-\infty,0)\times (0,t)$ in the left-half plane (see Figure~\ref{fig:LHP_RHP}) we find

\begin{align}\nonumber
&&\int_{0}^t\int_{-\infty}^0 (e^{-ikx+{(\sigma_Lk)^2} s}v^L(x,t))_s-(\sigma_L^2e^{-ikx+{(\sigma_L k)^2} s}(v^L_x(x,s)+ik v^L(x,s)))_x \ud x\ud s=0\\\la{globalleft}
&\Rightarrow\!\!\!\!\!&\int_{-\infty}^0\!\! e^{-ikx} v^L_0(x) \ud x-\int_{-\infty}^0\!\! e^{-ikx+{(\sigma_L k)^2} t}{v^L(x,t)}\ud x+\int_{0}^t \!\sigma_L^2 e^{{(\sigma_Lk)^2} s}(v^L_x(0,s)+i kv^L(0,s))\ud s=0.
\end{align}

\begin{figure}[htbp]
\begin{center}
\def\svgwidth{3in}
   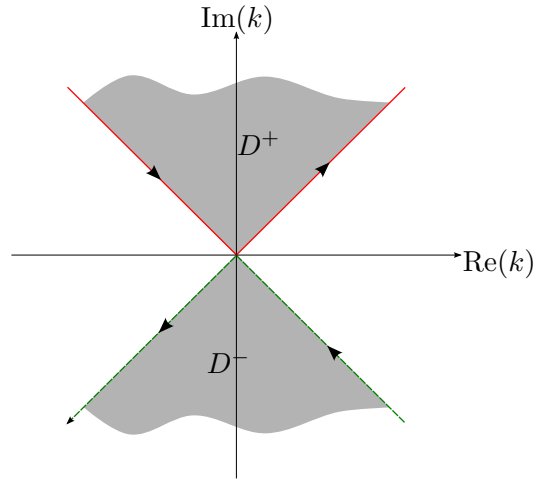 
   \caption{The Domains $D^+$ and $D^-$ for the heat equation.   \label{fig:heateqn}}
   \end{center}
\end{figure}

Let $\CC^+=\{z\in\CC:\Im(z)\geq0\}$.  Similarly, let $\CC^-=\{z\in\CC:\Im(z)\leq0\}$.  Since $|x|$ can become arbitrarily large, we require $k\in\CC^+$ in \rf{globalleft} in order to guarantee that the first two integrals are well defined. Let $D=\{k\in\CC:\Re({k^2})<0\}=D^+\cup D^-$.  The region $D$ is shown in Figure~\ref{fig:heateqn}.

For $k\in\CC$ we define the following transforms:

\begin{align*}
g_{0}({\omega},t)&=\int_{0}^te^{\omega s}v^L(0,s)\ud s=\int_{0}^te^{\omega s}(v^R(0,s)+\gamma^R-\gamma^L)\ud s\\
&=\frac{(\gamma^R-\gamma^L)(e^{\omega t}-1)}{\omega}+\int_{0}^te^{\omega s}v^R(0,s)\ud s,\\
g_{1}({\omega},t)&=\int_{0}^te^{{\omega} s}v^L_x(0,s)\ud s=\frac{\sigma_R^2}{\sigma_L^2} \int_{0}^te^{{\omega} s}v^R_x(0,s)\ud s,\\
\hat{v}^L(k,t)&=\int_{-\infty}^0e^{-ikx}v^L(x,t)\ud x, &\hat{v}^L_0(k)&=\int_{-\infty}^0e^{-ikx}v^L_0(x)\ud x,\\
\hat{v}^R(k,t)&=\int_{0}^\infty e^{-ikx}v^R(x,t)\ud x, &\hat{v}^R_0(k)&=\int_{0}^\infty e^{-ikx}v^R_0(x)\ud x.
\end{align*}

\no Using these definitions, the global relation \rf{globalleft} is rewritten as
\beq\label{GRLnz}
\hat{v}^L_0(k)-e^{(\sigma_Lk)^2 t}\hat{v}^L(k,t)+ik\sigma_L^2g_{0}((\sigma_Lk)^2,t)+\sigma_L^2g_{1}((\sigma_Lk)^2,t)=0,~~~k\in\CC^+.
\eeq

Since the dispersion relation $\omega_L(k)=(\sigma_L k)^2$ is invariant under $k\to -k$, so are $g_0((\sigma_Lk)^2,t)$ and $g_1((\sigma_Lk)^2,t)$. Thus we can supplement \eqref{GRLnz} with its evaluation at $-k$, namely

\beq\label{GRLnz minus}
\hat{v}^L_0(-k)-e^{(\sigma_Lk)^2 t}\hat{v}^L(-k,t)-ik\sigma_L^2g_{0}((\sigma_Lk)^2,t)+\sigma_L^2g_{1}((\sigma_Lk)^2,t)=0.
\eeq

\no This relation is valid on $k\in\CC^-$.  Using Green's Formula on $(0,\infty)\times (0,t)$ (see Figure~\ref{fig:LHP_RHP}), the global relation for $v^R(x,t)$ is

\beq\label{GRRnz}
\hat{v}^R_0(k)-e^{{(\sigma_R k)^2} t}\hat{v}^R(k,t)-ik\sigma_R^2\left(\!\!g_{0}((\sigma_R k)^2,t)\!+\!\frac{(\gamma^L-\gamma^R)(e^{(\sigma_R k)^2 t}-1)}{(\sigma_R k)^2}\!\!\right)-\sigma_L^2g_{1}((\sigma_R k)^2,t)\!=\!0,
\eeq

\no valid in $k\in\CC^-$.

As above, using the invariance of $\omega_R(k)=(\sigma_R k)^2$, $g_0((\sigma_R k)^2,t)$, and $g_1((\sigma_R k)^2,t)$ under $k\to-k$, we supplement \eqref{GRRnz} with

\beq\label{GRRnz minus}
\begin{split}
\hat{v}^R_0(-k)-e^{{(\sigma_R k)^2} t}\hat{v}^R(-k,t)+&ik\sigma_R^2\left(g_{0}((\sigma_R k)^2,t)+\frac{(\gamma^L-\gamma^R)(e^{(\sigma_R k)^2 t}-1)}{(\sigma_R k)^2}\right)\\
&~~~~~~~~~~~~~~~~~~~~~~-\sigma_L^2g_{1}((\sigma_R k)^2,t)=0,
\end{split}
\eeq	

\no for $k\in\CC^+$.

Inverting the Fourier transforms in \eqref{GRLnz} we have

\beq\label{solnL1nz}
v^L(x,t)=\frac{1}{2\pi}\int_{-\infty}^\infty e^{ikx-{(\sigma_Lk)^2} t}\hat{v}^L_0(k)\ud k+\frac{\sigma_L^2}{2\pi}\int_{-\infty}^\infty e^{ikx-(\sigma_Lk)^2 t}(ikg_0((\sigma_Lk)^2,t)+g_{1}((\sigma_Lk)^2,t))\ud k
\eeq

\no for $x\in(-\infty,0)$ and $t>0$.  The integrand of the second integral in \eqref{solnL1nz} is entire and decays as $k\to\infty$ for $k\in\CC^-\setminus D^-$.  Using the analyticity of the integrand and applying Jordan's Lemma~\cite{AblowitzFokas} we can replace the contour of integration of the second integral by $-\int_{\partial D^-}$:

\beq\label{solnLnz}
v^L(x,t)=\frac{1}{2\pi}\int_{-\infty}^\infty e^{ikx-{(\sigma_Lk)^2} t}\hat{v}^L_0(k)\ud k-\frac{\sigma_L^2}{2\pi}\int_{\partial D^-} e^{ikx-(\sigma_Lk)^2 t}(ikg_0((\sigma_Lk)^2,t)+g_{1}((\sigma_Lk)^2,t))\ud k.
\eeq

Proceeding similarly on the right, starting from \eqref{GRRnz}, we have

\begin{align}
v^R(x,t)&=\frac{1}{2\pi}\int_{-\infty}^\infty e^{ikx-(\sigma_R k)^2 t}\hat{v}^R_0(k)\ud k-\nonumber\\
&~~-\frac{1}{2\pi}\int_{-\infty}^\infty e^{ikx-(\sigma_R k)^2 t}\left(ik\sigma_R^2\left(g_{0}((\sigma_R k)^2,t)+\frac{(\gamma^L-\gamma^R)(e^{(\sigma_R k)^2 t}-1)}{(\sigma_R k)^2}\right)+\sigma_L^2g_{1}((\sigma_R k)^2,t)\right)\ud k,\nonumber\\
&=\frac{\gamma^L-\gamma^R}{2}\left(1-\erf\left(\frac{x}{2\sqrt{\sigma_R^2 t}}\right)\right)+\frac{1}{2\pi}\int_{-\infty}^\infty e^{ikx-(\sigma_R k)^2 t}\hat{v}^R_0(k)\ud k \nonumber\\
&~~-\frac{1}{2\pi}\int_{-\infty}^\infty e^{ikx-(\sigma_R k)^2 t}(ik\sigma_R^2g_{0}((\sigma_R k)^2,t)+\sigma_L^2g_{1}((\sigma_R k)^2,t))\ud k,\nonumber\\
&=\frac{\gamma^L-\gamma^R}{2}\left(1-\erf\left(\frac{x}{2\sqrt{\sigma_R^2 t}}\right)\right)+\frac{1}{2\pi}\int_{-\infty}^\infty e^{ikx-(\sigma_R k)^2 t}\hat{v}^R_0(k)\ud k \nonumber\\
&~~-\frac{1}{2\pi}\int_{\partial D^+} e^{ikx-(\sigma_R k)^2 t}(ik\sigma_R^2g_{0}((\sigma_R k)^2,t)+\sigma_L^2g_{1}((\sigma_R k)^2,t))\ud k.\label{solnR1nz}
\end{align}

\no for $x\in(0,\infty)$ and $t>0$. Here $\erf(\cdot)$ denotes the error function: $\erf(z)=\frac{2}{\sqrt{\pi}}\int_{0}^z \exp(-y^2)\ud y$. To obtain the second equality above we integrated the terms that are explicit.

The expressions \rf{solnLnz} and \rf{solnR1nz} for $v^L(x,t)$ and $v^R(x,t)$ depend on the unknown functions $g_0$ and $g_1$, evaluated at different arguments. These functions need to be expressed in terms of known quantities.  To obtain a system of two equations for the two unknown functions we use \eqref{GRLnz minus} and \eqref{GRRnz} for $g_0((\sigma_L k)^2,t)$, and $g_1((\sigma_L k)^2,t)$. This requires the transformation $k\to -\sigma_L k/\sigma_R$ in \eqref{GRRnz}. The $-$ sign is required to ensure that both equations are valid on $\mathbb{C}^-$, allowing for their simultaneous solution. We find

\begin{align}\nonumber
g_{0}((\sigma_Lk)^2,t)&=\frac{i}{k\sigma_L(\sigma_L+\sigma_R)}(e^{(\sigma_Lk)^2 t}(\hat{v}^L(-k,t)+\hat{v}^R(k\sigma_L/\sigma_R,t))-\hat{v}^L_0(-k)-\hat{v}^R_0(k\sigma_L/\sigma_R))\\\la{whoa1}
&+\frac{(\gamma_L-\gamma_R)(1-e^{(\sigma_Lk)^2t})}{(\sigma_Lk)^2(\sigma_L+\sigma_R)},\\\nonumber
g_{1}((\sigma_Lk)^2,t)&=\frac{1}{\sigma_L^2(\sigma_L+\sigma_R)}(e^{(\sigma_Lk)^2 t}(\sigma_R\hat{v}^L(-k,t)-\sigma_L\hat{v}^R(k\sigma_L/\sigma_R,t))+\sigma_L\hat{v}^R_0(k\sigma_L/\sigma_R)\\\la{whoa2}
&-\sigma_R\hat{v}^L_0(-k))+\frac{i(\gamma_L-\gamma_R)(1-e^{(\sigma_Lk)^2t})}{k\sigma_L^2(\sigma_L+\sigma_R)},
\end{align}

\no valid for $k\in \CC^-$. These expressions are substituted into \eqref{solnLnz} and \eqref{solnR1nz}. This results in expressions for $v^L(x,t)$ and $v^R(x,t)$ that appear to depend on $v^L$ and $v^R$ themselves. We examine the contribution of the terms involving $\hat{v}^L$ and $\hat{v}^R$. Starting with \eqref{solnLnz} we obtain for $v^L(x,t)$ the following expression:

\begin{align}
v^L(x,t)&=\frac{\sigma_R(\gamma^R-\gamma^L)}{\sigma_L+\sigma_R}\left(1+\erf\left(\frac{x}{2\sqrt{\sigma_L^2t}}\right)\right)+\frac{1}{2\pi}\int_{-\infty}^\infty e^{ikx-{(\sigma_Lk)^2} t}\hat{v}^L_0(k)\ud k \nonumber\\
&+\int_{\partial D^-}\frac{\sigma_R-\sigma_L}{2\pi(\sigma_L+\sigma_R)}e^{ikx-(\sigma_Lk)^2t}\hat{v}^L_0(-k) \ud k-\int_{\partial D^-}\frac{\sigma_L}{\pi(\sigma_L+\sigma_R)}e^{ikx-(\sigma_Lk)^2t}\hat{v}^R_0(k\sigma_L/\sigma_R)\ud k \nonumber\\
&+\int_{\partial D^-}\frac{\sigma_L-\sigma_R}{2\pi(\sigma_L+\sigma_R)}e^{ikx}\hat{v}^L(-k,t) \ud k+\int_{\partial D^-}\frac{\sigma_L}{\pi(\sigma_L+\sigma_R)}e^{ikx}\hat{v}^R(k\sigma_L/\sigma_R,t)\ud k,\label{fullsolnL}
\end{align}

\begin{figure}[tb]
   \centering
\def\svgwidth{3in}
   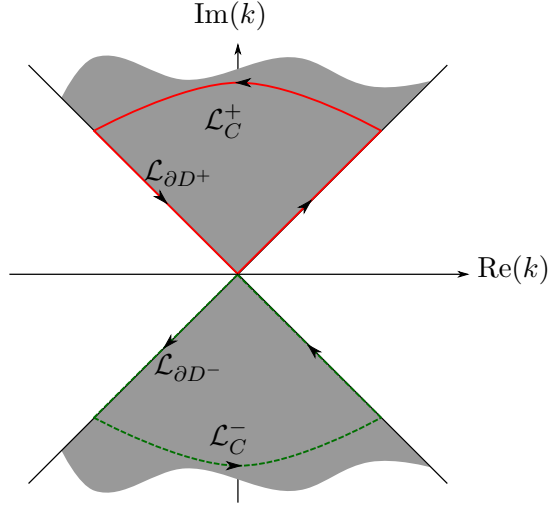 
   \caption{The countour $\mathcal{L}^-$ is shown in green.  An application of Cauchy's Integral Theorem ~\cite{AblowitzFokas} using this contour allows elimination of the contribution of $\hat{v}^L(-k,t)$ from the integral expression~\eqref{fullsolnL}.  Similarly, the contour $\mathcal{L^+}$ is shown in red and application of Cauchy's Integral Theorem using this contour allows elimination of the contribution of $\hat{v}^R(-k,t)$ from the integral expression~\eqref{fullsolnR}.   \label{fig:LHP_UHPclose}}
\end{figure}

\no for $x\in(-\infty,0)$, $t>0$.  The first four terms depend only on known functions.  To examine the second-to-last term we notice that the integrand is analytic for all $k\in \CC^-$ and that $\hat{v}^L(-k,t)$ decays for $k\to\infty$ for $k\in\CC^-$.  Thus, by Jordan's Lemma, the integral of $\exp(ikx)\hat{v}^L(-k,t)$ along a closed, bounded curve in $\CC^-$ vanishes. In particular we consider the closed curve $\mathcal{L}^-=\mathcal{L}_{\partial D^-}\cup\mathcal{L}^-_C$ where $\mathcal{L}_{\partial D^-}=\partial D^- \cap \{k: |k|<C\}$ and $\mathcal{L}^-_C=\{k\in D^-: |k|=C\}$, see Figure~\ref{fig:LHP_UHPclose}.

Since the integral along $\mathcal{L}_C^-$ vanishes for large $C$, the fourth integral on the right-hand side of \eqref{fullsolnL} must vanish since the contour $\mathcal{L}_{\partial D^-}$  becomes $\partial D^-$ as $C\to\infty$.  The uniform decay of $\hat{v}^L(-k,t)$ for large $k$ is exactly the condition required for the integral to vanish, using Jordan's Lemma.  For the final integral in Equation~\eqref{fullsolnL} we use the fact that $\hat{v}^R(k \sigma_L/\sigma_R,t)$ is analytic and bounded for $k\in\CC^-$.  Using the same argument as above, the fifth integral in~\eqref{fullsolnL} vanishes and we have an explicit representation for $v^L(x,t)$ in terms of initial conditions:

\begin{align}
v^L(x,t)&=\frac{\sigma_R(\gamma^R-\gamma^L)}{\sigma_L+\sigma_R}\left(1+\erf\left(\frac{x}{2\sqrt{\sigma_L^2t}}\right)\right)+\frac{1}{2\pi}\int_{-\infty}^\infty e^{ikx-{(\sigma_Lk)^2} t}\hat{v}^L_0(k)\ud k \nonumber\\
&+\int_{\partial D^-}\frac{\sigma_R-\sigma_L}{2\pi(\sigma_L+\sigma_R)}e^{ikx-(\sigma_Lk)^2t}\hat{v}^L_0(-k) \ud k-\int_{\partial D^-}\frac{\sigma_L}{\pi(\sigma_L+\sigma_R)}e^{ikx-(\sigma_Lk)^2t}\hat{v}^R_0(k\sigma_L/\sigma_R)\ud k.\label{TOTALsolnL}
\end{align}

To find an explicit expression for $v^R(x,t)$ we need to evaluate $g_0$ and $g_1$ at different arguments, also ensuring that the expressions are valid for $k\in \CC^+\setminus D^+$.  From \rf{whoa1} and \rf{whoa2}, we find

\bea
g_{0}((\sigma_R k)^2,t)&=&\frac{-i}{k\sigma_R(\sigma_L+\sigma_R)}(e^{(\sigma_Rk)^2 t}(\hat{v}^L(k\sigma_R/\sigma_L,t)+\hat{v}^R(-k,t))-\hat{v}^L_0(k\sigma_R/\sigma_L)-\hat{v}^R_0(-k))\\
&&+\frac{(\gamma^L-\gamma^R)(1-e^{(\sigma_Rk)^2t})}{k^2\sigma_R(\sigma_L+\sigma_R)},\\
g_{1}((\sigma_R k)^2,t)&=&\frac{1}{\sigma_L^2(\sigma_L+\sigma_R)}(e^{(\sigma_Rk)^2 t}(\sigma_R\hat{v}^L(k\sigma_R/\sigma_L,t)-\sigma_L\hat{v}^R(-k,t))+\sigma_L\hat{v}^R_0(-k)\\
&&-\sigma_R\hat{v}^L_0(k\sigma_R/\sigma_L))-\frac{i(\gamma^L-\gamma^R)(1-e^{(\sigma_Rk)^2t})}{k\sigma_L^2(\sigma_L+\sigma_R)}.
\eea

\no Substituting these into equation~\eqref{solnR1nz}, we obtain

\begin{align}
v^R(x,t)&=\frac{\sigma_L(\gamma^L-\gamma^R)}{\sigma_L+\sigma_R}\left(1-\erf\left(\frac{x}{2\sqrt{\sigma_R^2 t}}\right)\right)+\frac{1}{2\pi}\int_{-\infty}^\infty e^{ikx-(\sigma_R k)^2 t}\hat{v}^R_0(k)\ud k \nonumber\\
&+\int_{\partial D^+}\frac{\sigma_R-\sigma_L}{2\pi(\sigma_L+\sigma_R)}e^{ikx-(\sigma_Rk)^2t}\hat{v}^R_0(-k) \ud k+\int_{\partial D^+}\frac{\sigma_R}{\pi(\sigma_L+\sigma_R)}e^{ikx-(\sigma_Rk)^2t}\hat{v}^L_0(k\sigma_R/\sigma_L,t)  \ud k \nonumber\\
&+\int_{\partial D^+}\frac{\sigma_L-\sigma_R}{2\pi(\sigma_L+\sigma_R)}e^{ikx}\hat{v}^R(-k,t) \ud k-\int_{\partial D^+}\frac{\sigma_R}{\pi(\sigma_L+\sigma_R)}e^{ikx}\hat{v}^L(k\sigma_R/\sigma_L,t)  \ud k.\label{fullsolnR}
\end{align}

\no for $x\in(0,\infty)$, $t>0$.  As before, everything about the first three integrals is known.  To compute the fourth integral we proceed as we did before for $v^L(x,t)$ and eliminate integrals that decay in the regions over which we are integrating.  The final solution is

\begin{align}
v^R(x,t)&=\frac{\sigma_L(\gamma^L-\gamma^R)}{\sigma_L+\sigma_R}\left(1-\erf\left(\frac{x}{2\sqrt{\sigma_R^2 t}}\right)\right)+\frac{1}{2\pi}\int_{-\infty}^\infty e^{ikx-(\sigma_R k)^2 t}\hat{v}^R_0(k)\ud k \nonumber\\
&+\int_{\partial D^+}\frac{\sigma_R-\sigma_L}{2\pi(\sigma_L+\sigma_R)}e^{ikx-(\sigma_Rk)^2t}v^R_0(-k) \ud k+\int_{\partial D^+}\frac{\sigma_R}{\pi(\sigma_L+\sigma_R)}e^{ikx-(\sigma_Rk)^2t}v^L_0(k\sigma_R/\sigma_L)  \ud k.\label{TOTALsolnR}
\end{align}

Returning to the original variables we have the following proposition which determines $u^R$ and $u^L$ fully explicitly in terms of the given initial conditions and the prescribed boundary conditions as $|x|\ra\infty$.

\begin{prop}\la{2infprop}
The solution of the heat transfer problem (\ref{ueqns})-(\ref{bcs}) is given by
\begin{subequations}\la{usolns}
\begin{align}
u^L(x,t)&=\gamma^L+\frac{\sigma_R(\gamma^R-\gamma^L)}{\sigma_L+\sigma_R}\left(1-\erf\left(\frac{x}{2\sqrt{\sigma_L^2t}}\right)\right)+\frac{1}{2\pi}\int_{-\infty}^\infty e^{ikx-{(\sigma_Lk)^2} t}\hat{v}^L_0(k)\ud k \nonumber\\
&+\int_{\partial D^-}\frac{\sigma_R-\sigma_L}{2\pi(\sigma_L+\sigma_R)}e^{ikx-(\sigma_Lk)^2t}\hat{v}^L_0(-k) \ud k-\int_{\partial D^-}\frac{\sigma_L}{\pi(\sigma_L+\sigma_R)}e^{ikx-(\sigma_Lk)^2t}\hat{v}^R_0(k\sigma_L/\sigma_R)\ud k,\label{TOTALsoln2iL}\\
u^R(x,t)&=\gamma^R+\frac{\sigma_L(\gamma^L-\gamma^R)}{\sigma_L+\sigma_R}\left(1-\erf\left(\frac{x}{2\sqrt{\sigma_R^2 t}}\right)\right)+\frac{1}{2\pi}\int_{-\infty}^\infty e^{ikx-(\sigma_R k)^2 t}\hat{v}^R_0(k)\ud k \nonumber\\
&+\int_{\partial D^+}\frac{\sigma_R-\sigma_L}{2\pi(\sigma_L+\sigma_R)}e^{ikx-(\sigma_Rk)^2t}v^R_0(-k) \ud k+\int_{\partial D^+}\frac{\sigma_R}{\pi(\sigma_L+\sigma_R)}e^{ikx-(\sigma_Rk)^2t}v^L_0(k\sigma_R/\sigma_L)  \ud k.\label{TOTALsoln2iR}
\end{align}
\end{subequations}
\end{prop}
\vspace*{0.1in}

{\bf Remarks.}

\begin{itemize}

\item The use of the discrete symmetries of the dispersion relation is an important aspect of the Fokas Method \cite{DeconinckTrogdonVasan, FokasBook, FokasPelloni4}. When solving the heat equation in a single medium, the only discrete symmetry required is $k\to -k$, which was used here as well to obtain \rf{GRLnz minus} and \rf{GRRnz minus}. Due to the two media, there are two dispersion relations in the present problem: $\omega_1=(\sigma_L k)^2$ and $\omega_2=(\sigma_R k)^2$. The collection of both dispersion relations $\{\omega_1, \omega_2\}$ retains the discrete symmetry $k\to -k$, but admits two additional ones, namely: $k\to  (\sigma_R/\sigma_L)k$ and $k\to  (\sigma_L/\sigma_R)k$, which transform the two dispersion relations to each other.  All nontrivial discrete symmetries of $\{\omega_1, \omega_2\}$ are needed to derive the final solution representation, and indeed they are used \emph{e.g.} to obtain the relations \rf{whoa1} and \rf{whoa2}.

\item With $\sigma_L=\sigma_R$ and $\gamma^L=\gamma^R=0$,  the solution formulae \eqref{usolns} in their proper $x$-domain of definition reduce to
the solution of the whole line problem as given in~\cite{FokasBook}.

\item Classical approaches to the problem presented in this section can be found in the literature, for the case $\gamma_L=0=\gamma_R$. For instance, for one special pair of initial conditions, a solution is presented in \cite{HahnO}. No explicit solution formulae using classical methods with general initial conditions exist to our knowledge. At best, one is left with having to find the solution of an equation involving inverse Laplace transforms, where the unknowns are embedded within these inverse transforms.

\item \sloppypar The steady-state solution to \eqref{ueqns} with initial conditions which decay sufficiently fast to the boundary values \eqref{ubcs} at $\pm \infty$ is easily obtained by letting $t\rightarrow \infty$ in~\eqref{usolns}.  This gives $\lim_{t\rightarrow \infty} u^R(x,t)=\lim_{t\rightarrow \infty} u^L(x,t)=(\gamma^L\sigma_L+\gamma^R\sigma_R)/(\sigma_L+\sigma_R)$. This is the weighted average of the boundary conditions at infinity with weights given by $\sigma_L$ and $\sigma_R$. To our knowledge this is a new result. It is consistent with the steady state limit $(\gamma^L+\gamma^R)/2$ for the whole-line problem with initial conditions that limit to different values $\gamma^L$ and $\gamma^R$ as $x\rightarrow \pm \infty$. This result is easily obtained from the solution of the heat equation defined on the whole line using the Fokas Method, but it can also be observed by employing piecewise-constant initial data in the classical Green's function solution, as described in Theorem 4-1 on page 171 (and comments thereafter) of~\cite{GuentherLee}. It should be emphasized that the steady state problem for (\ref{ueqns}-\ref{bcs}) or even for the heat equation defined on the whole line with different boundary conditions at $+\infty$ and $-\infty$ is ill posed in the sense that the steady solution cannot satisfy the boundary conditions.

\end{itemize}

Using a slight variation on the method presented in~\cite{FlyerFokas} one can compute the solutions \eqref{usolns} numerically with specified initial conditions.  We plot solutions for the case of vanishing boundary conditions ($\gamma^L=\gamma^R=0$) with
\bea
u^L_0(x)&=&x^2e^{\alpha_L^2 x},\\
u^R_0(x)&=&x^2e^{-\alpha_R^2 x},
\eea

\no with $\alpha_L=25$ and $\alpha_R=30$.  The Fourier transforms of these initial conditions may be computed explicitly.  We choose $\sigma_L=.02$ and $\sigma_R=.06$. The initial conditions are chosen so as to satisfy the interface boundary conditions (\ref{bcs}) at $t=0$. The results clearly illustrate the discontinuity in the first derivative of the temperature at the interface $x=0$.
\begin{figure}[htbp!] 
   \centering
   \includegraphics[width=.7\textwidth]{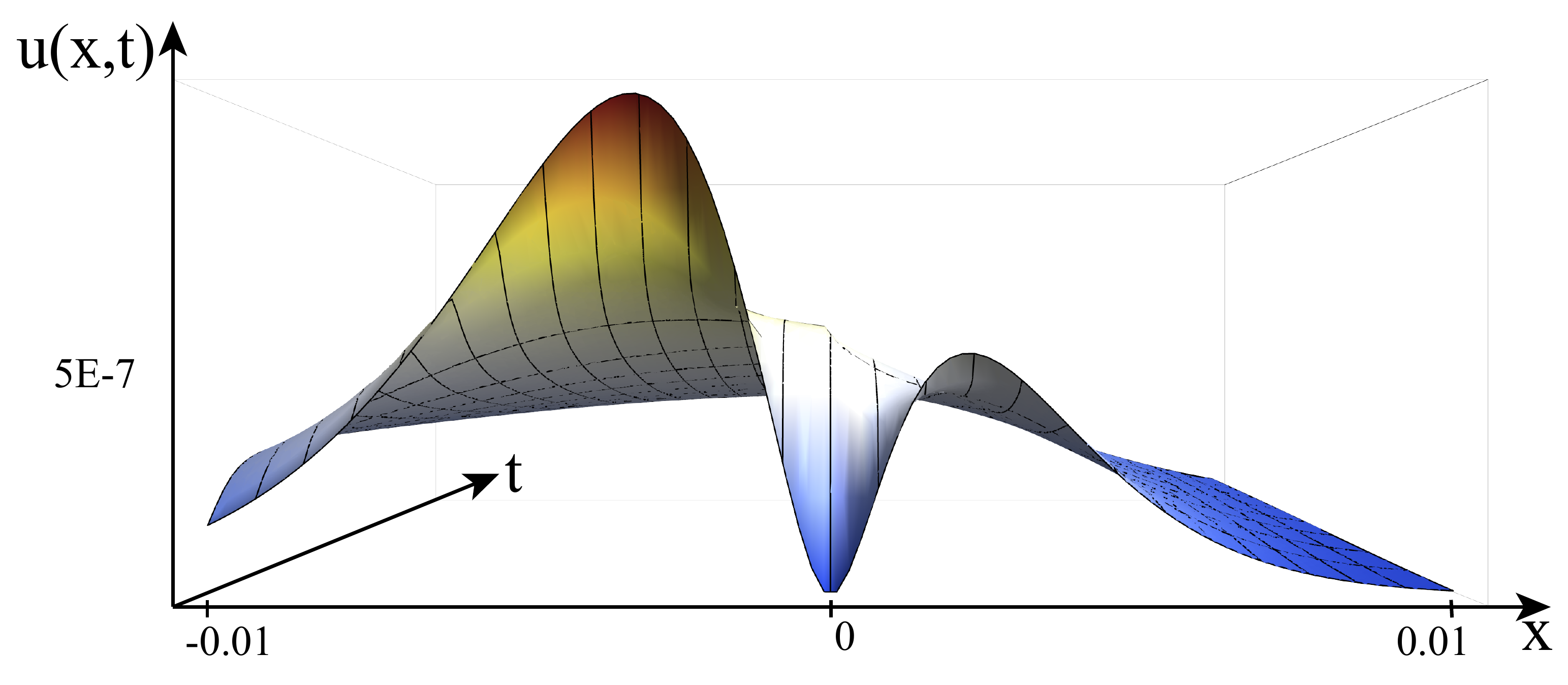}
   \includegraphics[width=.7\textwidth]{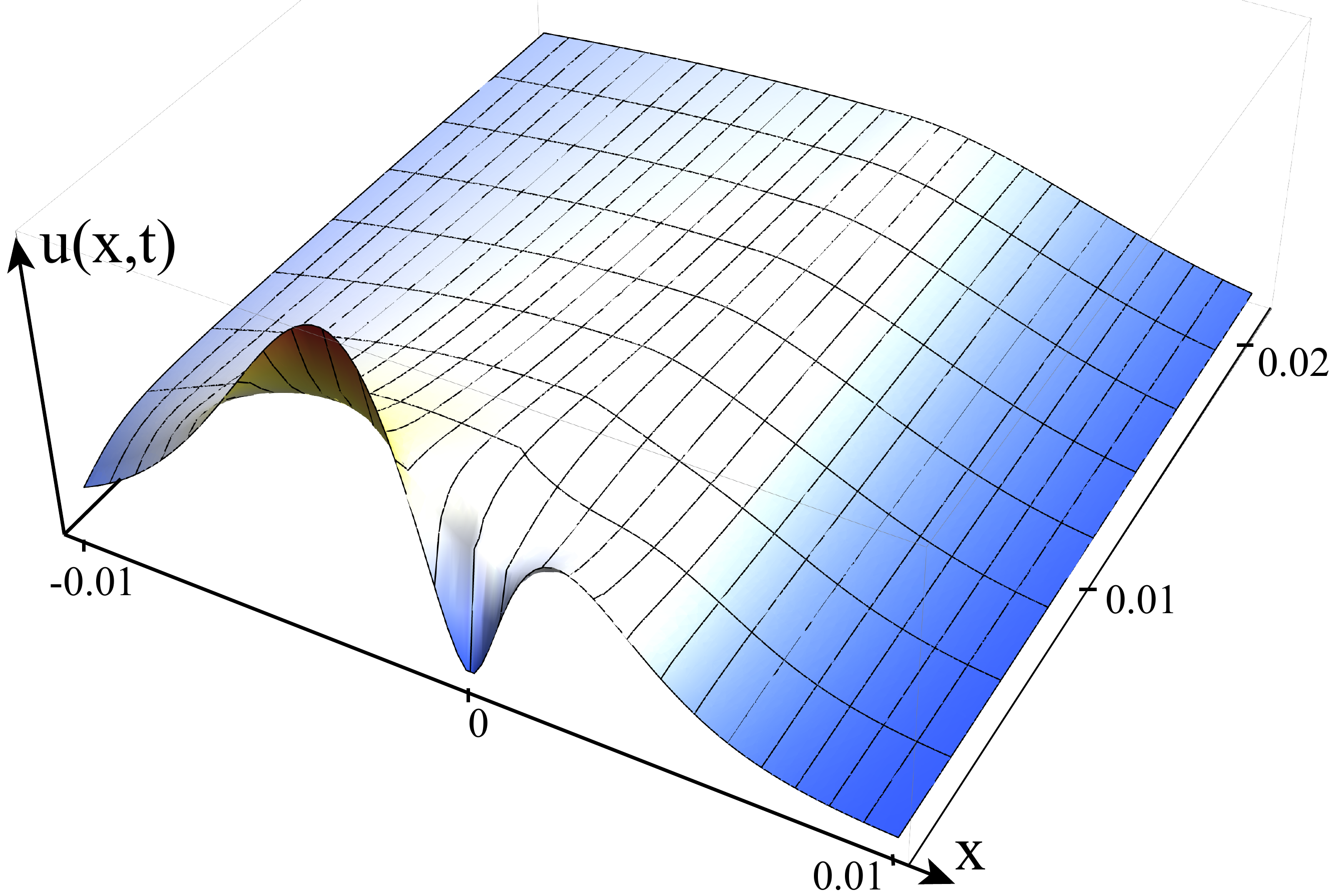}
      \includegraphics[width=.7\textwidth]{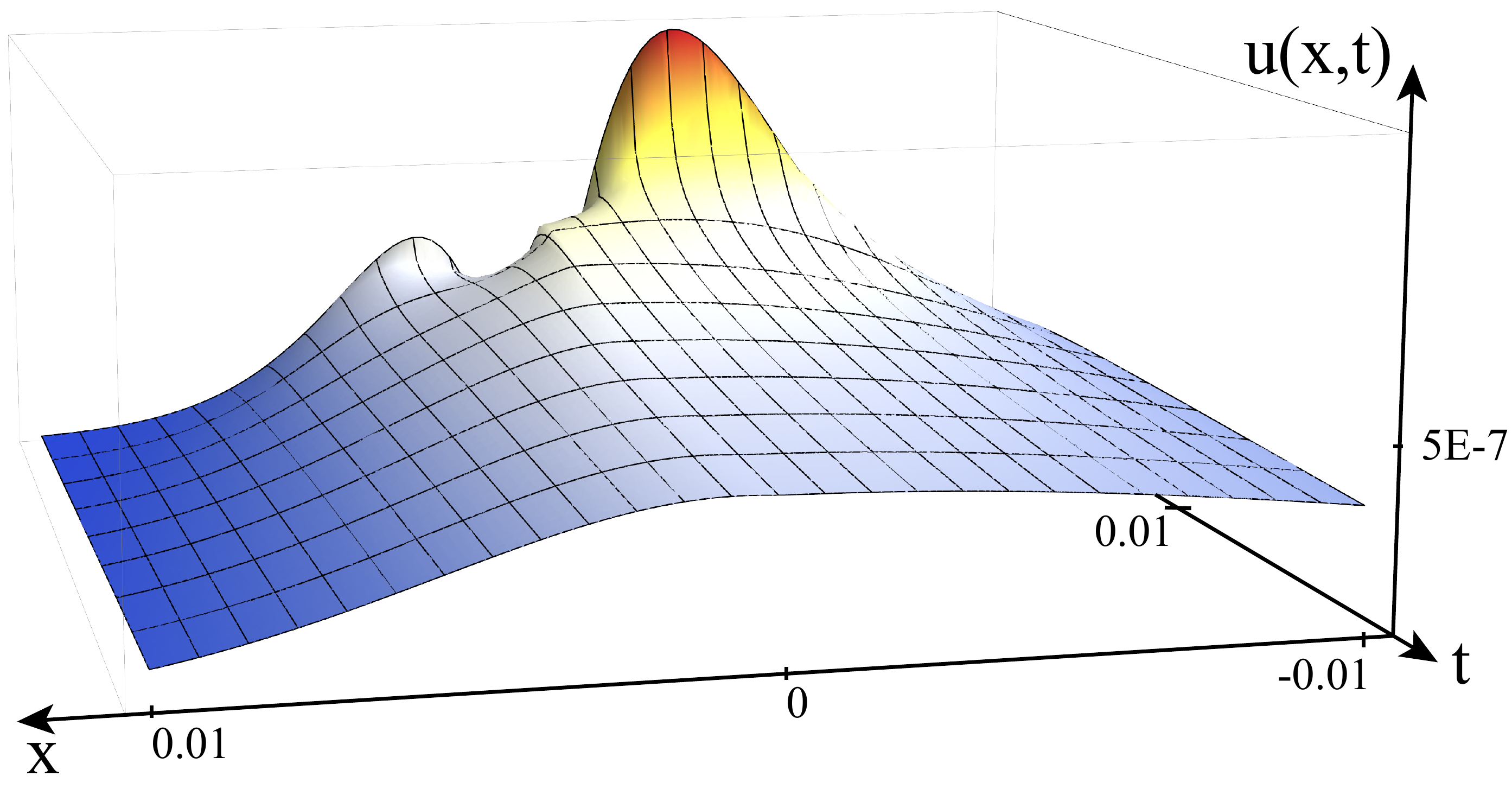}
   \caption{Results for the solution~\eqref{TOTALsolnL} and~\eqref{TOTALsolnR} with $u^L_0(x)=x^2e^{(25)^2 x}$, $u^R_0(x)=x^2e^{-(30)^2x}$ and $\sigma_L=.02$, $\sigma_R=.06$, $\gamma^L=\gamma^R=0$, $t\in[0,0.02]$ using the hybrid analytical-numerical method of \cite{FlyerFokas}.
   \label{fig:heat2i_num}}
\end{figure}

\section{Two finite domains}\la{sec:2f}

Next, we consider the problem of heat conduction through two walls of finite width (or of two finite rods) with Robin boundary conditions:

we seek two functions
$$
u^L(x,t),~\;x\in(-a,0),~ \;t\geq0,\qquad \qquad u^R(x,t),\;~~x\in(0,b),\;~~t\geq0,
$$

\no satisfying the equations

\begin{subequations}\label{doublelayerwall}
\begin{align}
u^L_t(x,t)&=\sigma_L^2 u^L_{xx}(x,t),&-a<x&<0,~t>0,\\
u^R_t(x,t)&=\sigma_R^2u^R_{xx}(x,t),&0<x&<b,~t>0,
\end{align}
\end{subequations}

\begin{subequations}
\no the initial conditions
\begin{align}
\label{2fics}
u^L(x,0)&=u^L_0(x),&-a<x&<0,\\
u^R(x,0)&=u^R_{0}(x),&0<x&<b,
\end{align}
\end{subequations}

\begin{subequations}
\no the boundary conditions
\begin{align}
\label{2fbcs}
f^L(t)&=\alpha_1 u^L(-a,t)+\alpha_2 u^L_x(-a,t),& t>0,\\
f^R(t)&=\alpha_3 u^R(b,t)+\alpha_4 u^R_x(b,t), & t>0,
\end{align}
\end{subequations}

\no and the continuity conditions
\begin{subequations}\label{bcs2}
\begin{align}
u^L(0,t)&=u^R(0,t), &t>0,\\
\sigma_L^2u^L_x(0,t)&=\sigma_R^2u^R_x(0,t), &t>0,
\end{align}
\end{subequations}

\begin{figure}[htbp]
   \centering
   \includegraphics[width=.7\textwidth]{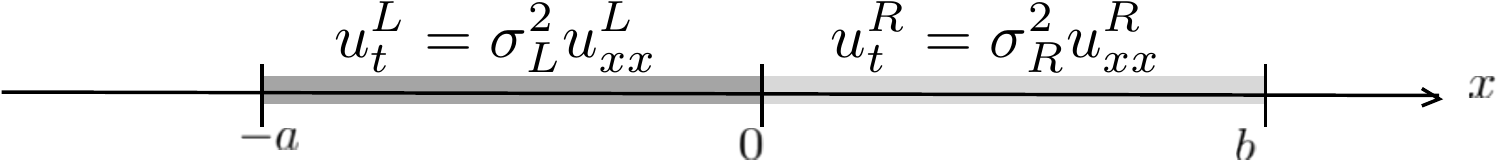} 
   \caption{The heat equation for two finite rods.   \label{fig:doublewall} }
\end{figure}

\no as illustrated in Figure~\ref{fig:doublewall}, where $a>0$, $b>0$ and $\alpha_i$, $1\leq i\leq 4$ constant.

If $\alpha_1=\alpha_3=0$ then Neumann boundary conditions are prescribed, whereas if $\alpha_2=\alpha_4=0$ then Dirichlet conditions are given.

As before we have the local relations

\begin{subequations}\label{local2f}
\begin{align}
&(e^{-ikx+(\sigma_Lk)^2 t}u^L(x,t))_t=(\sigma_L^2e^{-ikx+(\sigma_Lk)^2 t}(u^L_x(x,t)+ik u^L(x,t)))_x, \label{local2f_L}\\
&(e^{-ikx+(\sigma_R k)^2 t}u^R(x,t))_t=(\sigma_R^2 e^{-ikx+(\sigma_R k)^2 t}(u^R_x(x,t)+ik u^R(x,t)))_x.
\end{align}
\end{subequations}

We define the time transforms of the initial and boundary data and the spatial transforms of $u$ for $k\in\CC$ as follows:

\begin{align*}
&&\hat{u}^L_0(k)&=\int_{-a}^{0}e^{-ikx}u^L_0(x)\ud x,
&\hat{u}^L(k,t)&=\int_{-a}^{0}e^{-ikx}u^L(x,t)\ud x,\\
&&\hat{u}^R_0(k)&=\int_{0}^{b} e^{-ikx}u^R_0(x)\ud x,
&\hat{u}^R(k,t)&=\int_{0}^{b} e^{-ikx}u^R(x,t)\ud x,\\
&&\hat{f}_{L}(\omega,t)&=\int_{0}^te^{\omega s}{f}_L(s)\ud s,
&\hat{f}_{R}(\omega,t)&=\int_{0}^te^{\omega s}{f}_R(s)\ud s,\\
&& h_1^L(\omega,t)&=\int_{0}^te^{\omega s}u^L_x(-a,s)\ud s,
& h_0^L(\omega,t)&=\int_{0}^te^{\omega s}u^L(-a,s)\ud s,\\
&&h_1^R(\omega,t)&=\int_{0}^te^{\omega s}u^R_x(b,s)\ud s,
&h_0^R(\omega,t)&=\int_{0}^te^{\omega s}u^R(b,s)\ud s,\\
&&g_{1}(\omega,t)&=\int_{0}^te^{\omega s}u^L_x(0,s)\ud s=\frac{\sigma_R^2}{\sigma_L^2}\int_{0}^te^{\omega s}u^R_x(0,s)\ud s,\\
&&g_{0}(\omega,t)&=\int_{0}^te^{\omega s}u^L(0,s)\ud s=\int_{0}^te^{\omega s}u^R(0,s)\ud s.
\end{align*}

Using Green's Theorem on the domains $[-a,0]\times[0,t]$ and $[0,b]\times[0,t]$ respectively, we have the global relations

\begin{subequations}
\begin{align}\label{GL_dubwall}
e^{(\sigma_L k)^2 t}\hat{u}^L(k,t)&=\sigma_L^2(g_{1}((\sigma_Lk)^2,t)+ik g_{0}((\sigma_Lk)^2,t))- e^{ika}\sigma_L^2(h_1^L((\sigma_Lk)^2,t) \nonumber\\
&~~~~~~~~~~+ik h_0^L((\sigma_Lk)^2,t))+\hat{u}^L_0(k),\\
\label{GR_dubwall} e^{{(\sigma_R k)^2} t}\hat{u}^R(k,t)&=e^{-ikb}\sigma_R^2(h_1^R((\sigma_R k)^2,t)+ikh_0^R((\sigma_R k)^2),t)-
\sigma_L^2g_{1}((\sigma_R k)^2,t) \nonumber\\
&~~~~~~~~~~-ik\sigma_R^2 g_{0}((\sigma_R k)^2,t)+\hat{u}^R_0(k),
\end{align}
\end{subequations}

\no Both equations are valid for all $k\in\CC$, in contrast to~\eqref{GRLnz} and~\eqref{GRRnz}. Using the invariance of $\omega_L(k)=(\sigma_L k)^2$ and $\omega_R(k)=(\sigma_R k)^2$ under $k\to-k$ we obtain

\begin{subequations}
\begin{align}\nonumber
e^{(\sigma_Lk)^2 t}\hat{u}^L(-k,t)=&\sigma_L^2(g_{1}((\sigma_Lk)^2,t)-ik g_{0}((\sigma_Lk)^2,t))\\\label{neg_GL_dubwall} &~~~~~~~~~~-e^{-ika}\sigma_L^2(h_1^L((\sigma_Lk)^2,t)-ik h_0^L((\sigma_Lk)^2,t))+\hat{u}^L_0(-k),\\\nonumber
e^{{(\sigma_R k)^2} t}\hat{u}^R(-k,t)=&e^{ikb}\sigma_R^2(h_1^R((\sigma_R k)^2,t)-ikh_0^R((\sigma_R k)^2),t)\\\label{neg_GR_dubwall}
&~~~~~~~~~~-\sigma_L^2g_{1}((\sigma_R k)^2,t)+ik\sigma_R^2 g_{0}((\sigma_R k)^2,t)+\hat{u}^R_0(-k),
\end{align}
\end{subequations}

Inverting the Fourier transform in~\eqref{GL_dubwall},

\begin{align}\nonumber
u^L(x,t)=&\frac{1}{2\pi}\int_{-\infty}^\infty e^{ikx-{(\sigma_Lk)^2} t}\sigma_L^2(g_{1}((\sigma_Lk)^2,t)+ikg_{0}((\sigma_Lk)^2,t))\ud k\\\nonumber
&-\frac{1}{2\pi}\int_{-\infty}^\infty e^{ik(a+x)-{(\sigma_Lk)^2}t}\sigma_L^2(h_1^L((\sigma_Lk)^2,t)+ikh_0^L((\sigma_Lk)^2,t))\ud k\\\label{soln2L}
&+\frac{1}{2\pi}\int_{-\infty}^\infty e^{ikx-{(\sigma_Lk)^2} t}\hat{u}^L_0(k)\ud k.
\end{align}

\begin{figure}[htbp!] 
   \centering
   \def\svgwidth{5in}
   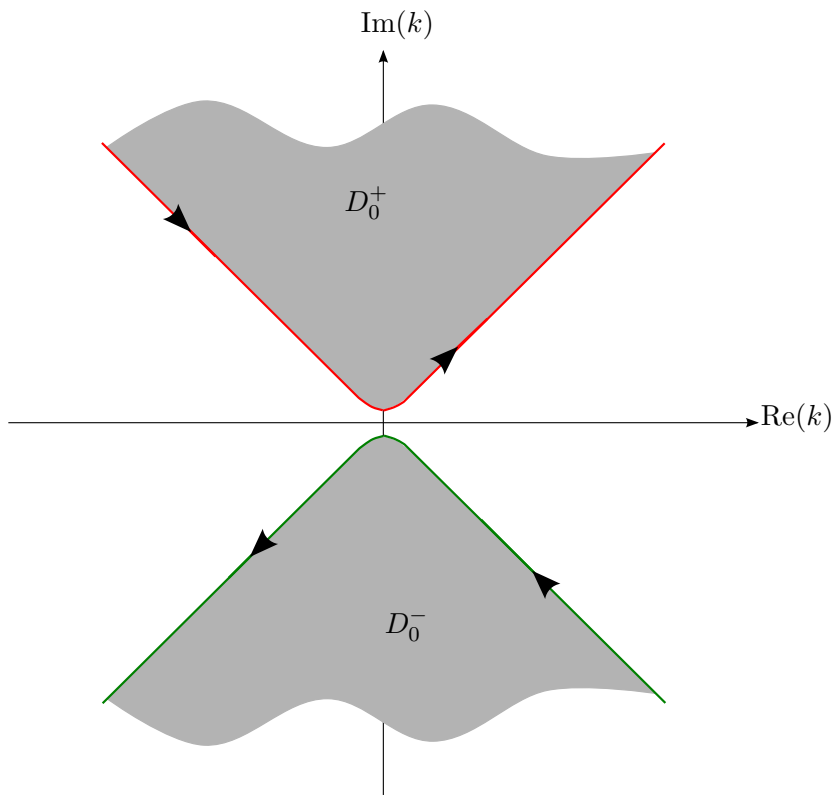
      \caption{Deformation of the contours in Figure~\ref{fig:heateqn} away from $k=0$.
   \label{fig:nonzero}}
\end{figure}

The integrand of the first integral is entire and decays as $k\to\infty$ for $k\in\CC^- \setminus D^-$.  The second integral has an integrand that is entire and decays as $k\to\infty$ for $k\in\CC^+\setminus D^+$.  It is convenient to deform both contours away from $k=0$ to avoid singularities in the integrands that become apparent in what follows. Initially, these singularities are removable, since the integrands are entire. Writing integrals of sums as sums of integrals, the singularities may cease to be removable. With the deformations away from $k=0$, the apparent singularities are no cause for concern.  In other words, we deform $D^+$ to $D^+_0$ and $D^-$ to $D^-_0$ as show in Figure~\ref{fig:nonzero}. Thus

\begin{align}\nonumber
u^L(x,t)=&\frac{-1}{2\pi}\int_{\partial D_0^-} e^{ikx-(\sigma_Lk)^2 t}\sigma_L^2(g_{1}((\sigma_Lk)^2,t)+ikg_{0}((\sigma_Lk)^2,t))\ud k\\
\nonumber
&-\frac{1}{2\pi}\int_{\partial D_0^+} e^{ik(a+x)-(\sigma_Lk)^2t}\sigma_L^2(h_1^L((\sigma_Lk)^2,t)+ikh_0^L((\sigma_Lk)^2,t))\ud k\\
\label{deformsoln2L}
&+\frac{1}{2\pi}\int_{-\infty}^\infty e^{ikx-(\sigma_Lk)^2 t}\hat{u}^L_0(k)\ud k.
\end{align}

To obtain the solution on the right we apply the inverse Fourier transform to~\eqref{GR_dubwall}:

\begin{align}\nonumber
u^R(x,t)=&\frac{1}{2\pi}\int_{-\infty}^\infty e^{ik(x-b)-{(\sigma_R k)^2} t}\sigma_R^2(h_1^R((\sigma_R k)^2,t)+ikh_0^R((\sigma_R k)^2,t))\ud k\\\nonumber
&-\frac{1}{2\pi}\int_{-\infty}^\infty e^{ikx-{(\sigma_R k)^2}t}(ik\sigma_R^2g_{0}((\sigma_R k)^2,t)+\sigma_L^2g_{1}((\sigma_R k)^2,t))\ud k\\\label{soln2R}
&+\frac{1}{2\pi}\int_{-\infty}^\infty e^{ikx-{(\sigma_R k)^2}t}\hat{u}^R_0(k)\ud k.
\end{align}

The integrand of the first integral is entire and decays as $k\to\infty$ for $k\in\CC^- \setminus D^-$.  The second integral has an integrand that is entire and decays as $k\to\infty$ for $k\in\CC^+\setminus D^+$.  We deform the contours as above to obtain

\begin{align}\nonumber
u^R(x,t)=&\frac{-1}{2\pi}\int_{\partial D_0^-} e^{ik(x-b)-{(\sigma_R k)^2} t}\sigma_R^2(h_1^R((\sigma_R k)^2,t)+ikh_0^R((\sigma_R k)^2,t))\ud k\\\nonumber
&-\frac{1}{2\pi}\int_{\partial D_0^+} e^{ikx-{(\sigma_R k)^2}t}(ik\sigma_R^2g_{0}((\sigma_R k)^2,t)+\sigma_L^2g_{1}((\sigma_R k)^2,t))\ud k\\\label{deformsoln2R}
&+\frac{1}{2\pi}\int_{-\infty}^\infty e^{ikx-{(\sigma_R k)^2}t}\hat{u}^R_0(k)\ud k.
\end{align}

Taking the time transform of the boundary conditions results in

$$
\hat{f}^L(\omega,s)=\alpha_1h_0^L(\omega,t)+\alpha_2h_1^L(\omega,t),
$$

\no and

$$
\hat{f}^R(\omega,s)=\alpha_3h_0^R(\omega,t)+\alpha_4h_1^R(\omega,t).
$$

These two equations together with~\eqref{neg_GL_dubwall} and~\eqref{neg_GR_dubwall} are four equations to be solved for the four unknowns $h_0^L(\omega, t)$, $h_0^R(\omega,t)$, $h_1^L(\omega, t)$, $h_1^R(\omega,t)$. The resulting expressions are substituted in \rf{deformsoln2L} and \rf{deformsoln2R}.

Although we could solve this problem in its full generality, we restrict to the case of Dirichlet boundary conditions ($\alpha_2=\alpha_4=0$), to simplify the already cumbersome formulae below. Then $h_0^L(\omega, t)$ and $h_0^R(\omega,t)$ are determined, and we solve two equations for two unknowns. The system \rf{neg_GL_dubwall}-\rf{neg_GR_dubwall} is not solvable for $h_1^L(\omega, t)$ and $h_1^R(\omega,t)$ if
$\Delta_L(k)=0$, where

\begin{align*}
\Delta_L(k)&=\pi(\sigma_L(e^{2iak}+1)(e^{2ibk\sigma_L/\sigma_R}-1)+\sigma_R(e^{2iak}-1)(e^{2ibk\sigma_L/\sigma_R}+1))\\
&=i \pi (e^{2iak}+1)(e^{2ibk\sigma_L/\sigma_R}+1)(\sigma_L \tan (bk\sigma_L/\sigma_R)+\sigma_R \tan(ak)).
\end{align*}

\no It is easily seen that all values of $k$ satisfying this (including $k=0$) are on the real line. Thus on the contours, the equations are solved without problem, resulting in the expressions below. As before, the right-hand sides of these expressions involve $\hat u^L(k,t)$ and $\hat u^L(k,t)$, evaluated at a variety of arguments. All terms with such dependence are written out explicitly below. Terms that depend on known quantities only are contained in $K^L$ and $K^R$, the expressions for which are given later.

\begin{align}\nonumber
u^L(x,t)&=K^L+\int_{\partial D_0^-} \frac{e^{ikx}(\sigma_L+\sigma_R)+e^{ik(x+2b\sigma_L/\sigma_R)}(\sigma_R-\sigma_L)}{2\Delta_L(k)}  \hat{u}^L(k,t)\ud k\\\nonumber
&+\int_{\partial D_0^-} \frac{-e^{ik(x+2a)}(\sigma_L+\sigma_R)+e^{ik(x+2a+2b\sigma_L/\sigma_R)}(\sigma_L-\sigma_R)}{2\Delta_L(k)}   \hat{u}^L(-k,t)     \ud k \\\nonumber
&+\int_{\partial D_0^-}  \frac{\sigma_L e^{ik(x+2a+2b\sigma_L/\sigma_R)}}{\Delta_L(k)}  \hat{u}^R(k\sigma_L/\sigma_R,t)     \ud k +
\int_{\partial D_0^-}  \frac{-\sigma_Le^{ik(x+2a)}}{\Delta_L(k)}  \hat{u}^R(-k\sigma_L/\sigma_R,t)   \ud k \\\nonumber
&+\int_{\partial D_0^+}  \frac{e^{ik(x+2a)}(\sigma_R-\sigma_L)+e^{ik(x+2a+2b\sigma_L/\sigma_R)}(\sigma_L+\sigma_R)}{2\Delta_L(k)}   \hat{u}^L(k,t)   \ud k \\\nonumber
&+\int_{\partial D_0^+}  \frac{-e^{ik(x+2a)}(\sigma_L+\sigma_R)+e^{ik(x+2a+2b\sigma_L/\sigma_R)}(\sigma_L-\sigma_R)}{2\Delta_L(k)}  \hat{u}^L(-k,t)   \ud k \\
&+\int_{\partial D_0^+}  \frac{\sigma_Le^{ik(x+2a+2b\sigma_L/\sigma_R)}}{\Delta_L(k)}  \hat{u}^R(k\sigma_L/\sigma_R,t)   \ud k +\int_{\partial D_0^+} \frac{-\sigma_L e^{ik(x+2a)}}{\Delta_L(k)}  \hat{u}^R(-k\sigma_L/\sigma_R,t)   \ud k ,\label{fullsoln2L}
\end{align}

\no and

\begin{align}\nonumber
u^R(x,t)&=K^R+\int_{\partial D_0^+}  \frac{ e^{ikx}\sigma_R }{\Delta_R(k)}  \hat{u}^L(k\sigma_R/\sigma_L,t)    \ud k +
\int_{\partial D_0^+}  \frac{ -\sigma_Re^{ik(x+2a\sigma_R/\sigma_L)}}{\Delta_R(k)}  \hat{u}^L(-k\sigma_R/\sigma_L,t)     \ud k \\\nonumber
&+\int_{\partial D_0^+}   \frac{e^{ik(2b+x)}(\sigma_L-\sigma_R)+e^{ik(x+2b+2a\sigma_R/\sigma_L)}(\sigma_L+\sigma_R)}{2\Delta_R(k)}  \hat{u}^R(k,t)     \ud k \\\nonumber
&+\int_{\partial D_0^+}   \frac{e^{ikx}(\sigma_R-\sigma_L)-e^{ik(x+2a\sigma_R/\sigma_L)}(\sigma_L+\sigma_R) }{2\Delta_R(k)}  \hat{u}^R(-k,t)   \ud k \\\nonumber
&+\int_{\partial D_0^-}   \frac{\sigma_R e^{ikx}}{\Delta_R(k)}   \hat{u}^L(k\sigma_R/\sigma_L,t)   \ud k +
\int_{\partial D_0^-}    \frac{-\sigma_Re^{ik(x+2a\sigma_R/\sigma_L)} }{\Delta_R(k)}    \hat{u}^L(-k\sigma_R/\sigma_L,t)   \ud k \\\nonumber
&+\int_{\partial D_0^-}   \frac{ e^{ikx}(\sigma_L+\sigma_R)-e^{ik(x+2a\sigma_R/\sigma_L)}(\sigma_L-\sigma_R) }{2\Delta_R(k)}   \hat{u}^R(k,t)   \ud k \\\label{fullsoln2R}
&+\int_{\partial D_0^-}  \frac{ e^{ikx}(\sigma_R-\sigma_L)-e^{ik(x+2a\sigma_R/\sigma_L)}(\sigma_L+\sigma_R)}{2\Delta_R(k)}  \hat{u}^R(-k,t)    \ud k,
\end{align}

\no where $\Delta_R(k)=\Delta_L(k\sigma_R/\sigma_L)$. The integrands written explicitly in \eqref{fullsoln2L} and \eqref{fullsoln2R} decay in the regions around whose boundaries they are integrated.  Thus, using Jordan's Lemma and Cauchy's Theorem these integrals are shown to vanish.  Thus the final solution is given by $K^L$ and $K^R$.

\begin{prop}\label{prop:2f} The solution of the heat transfer problem~\eqref{doublelayerwall}-\eqref{bcs2} is given by
\begin{align}\nonumber
&u^L(x,t)=K^L\\\nonumber
&=\int_{-\infty}^\infty e^{ikx-(\sigma_Lk)^2t}\hat{u}_0^L(k)\ud k+\int_{\partial D^-_0}\frac{-2ik\sigma_L^2\sigma_R e^{ik(x+2a+b\sigma_L/\sigma_R)-(\sigma_Lk)^2t}}{\alpha_3\Delta_L(k)}\hat{f}^R((\sigma_L k)^2,t)\ud k\\\nonumber
&+\int_{\partial D^-_0}\frac{ik\sigma_L^2e^{ik(x+a)-(\sigma_Lk)^2t}(\sigma_L+\sigma_R)-ik\sigma_L^2e^{ik(x+a+2b\sigma_L/\sigma_R)-(\sigma_Lk)^2t}(\sigma_L-\sigma_R)}{\alpha_1\Delta_L(k)}\hat{f}^L((\sigma_L k)^2,t)\ud k\\\nonumber
&+\int_{\partial D^-_0}\frac{-e^{ikx-(\sigma_Lk)^2t}(\sigma_L+\sigma_R)+e^{ik(x+2b\sigma_L/\sigma_R)-(\sigma_Lk)^2t}(\sigma_L-\sigma_R)}{2\Delta_L(k)}\hat{u}^L_0(k)\ud k\\\nonumber
&+\int_{\partial D^-_0}\frac{e^{ik(x+2a)-(\sigma_Lk)^2t}(\sigma_L+\sigma_R)+e^{ik(x+2a+2b\sigma_L/\sigma_R)-(\sigma_Lk)^2t}(\sigma_R-\sigma_L)}{2\Delta_L(k)}\hat{u}^L_0(-k)\ud k\\\nonumber
&+\int_{\partial D^-_0}\frac{-\sigma_Le^{ik(x+2a+2b\sigma_L/\sigma_R)-(\sigma_Lk)^2t}}{\Delta_L(k)}\hat{u}^R_0(k\sigma_L/\sigma_R)\ud k+
\int_{\partial D^-_0}\frac{\sigma_L e^{ik(x+2a)-(\sigma_Lk)^2t}}{\Delta_L(k)}\hat{u}^R_0(-k \sigma_L/\sigma_R)\ud k\\\nonumber
&+\int_{\partial D^+_0}\frac{ik\sigma_L^2e^{ik(x+a)-(\sigma_Lk)^2t}(\sigma_L+\sigma_R)-ik\sigma_L^2e^{ik(x+a+2b\sigma_L/\sigma_R)-(\sigma_Lk)^2t}(\sigma_L-\sigma_R)}{\alpha_1\Delta_L(k)}\hat{f}^L((\sigma_L k)^2,t)\ud k\\\nonumber
&+\int_{\partial D^+_0}\frac{-2ik\sigma_L^2\sigma_R e^{ik(x+2a+b\sigma_L/\sigma_R)-(\sigma_Lk)^2t}(1+\sigma_L\sigma_R)}{\alpha_3\Delta_L(k)}\hat{f}^R((\sigma_L k)^2,t)\ud k\\\nonumber
&+\int_{\partial D^+_0}\frac{-e^{ik(x+2a)-(\sigma_Lk)^2t}(\sigma_R-\sigma_L)-e^{ik(x+2a+2b\sigma_L/\sigma_R)-(\sigma_Lk)^2t}(\sigma_L+\sigma_R)}{2\Delta_L(k)}\hat{u}^L_0(k)\ud k\\\nonumber
&+\int_{\partial D^+_0}\frac{e^{ik(x+2a)-(\sigma_Lk)^2t}(\sigma_L+\sigma_R)+e^{ik(x+2a+2b\sigma_L/\sigma_R)-(\sigma_Lk)^2t}(\sigma_R-\sigma_L)}{2\Delta_L(k)}\hat{u}^L_0(-k)\ud k\\\label{1isolution}
&+\int_{\partial D^+_0}\frac{-\sigma_Le^{ik(x+2a+2b\sigma_L/\sigma_R)-(\sigma_Lk)^2t}}{\Delta_L(k)}\hat{u}^R_0(k\sigma_L/\sigma_R)\ud k+
\int_{\partial D^+_0}\frac{\sigma_Le^{ik(x+2a)-(\sigma_Lk)^2t}}{\Delta_L(k)}\hat{u}^R_0(-k\sigma_L/\sigma_R)\ud k,
\end{align}

\no for $-a<x<0$, and, for $0<x<b$

\begin{align}\nonumber
&u^R(x,t)=K^R\\\nonumber
&=\int_{-\infty}^\infty e^{ikx-(\sigma_Rk)^2t}\hat{u}_0^R(k)\ud k+
\int_{\partial D^-_0}\frac{2ik\sigma_L\sigma_R^2 e^{ik(x+a\sigma_R/\sigma_L)-(\sigma_Rk)^2t}}{\alpha_1\Delta_R(k)}\hat{f}^L((\sigma_R k)^2,t)\ud k\\\nonumber
&+\int_{\partial D^-_0}\frac{-ik\sigma_R^2e^{ik(x+b)-(\sigma_Rk)^2t}(\sigma_L-\sigma_R) -ik\sigma_R^2e^{ik(x+b+2a\sigma_R/\sigma_L)-(\sigma_Rk)^2t} }{\alpha_3\Delta_R(k)}\hat{f}^R((\sigma_R k)^2,t)\ud k\\\nonumber
&+\int_{\partial D^-_0}\frac{-\sigma_R e^{ikx-(\sigma_Rk)^2t}}{\Delta_R(k)}\hat{u}^L_0(k\sigma_R/\sigma_L)\ud k+\int_{\partial D^-_0}\frac{\sigma_R e^{ik(x+2a\sigma_R/\sigma_L)-(\sigma_Rk)^2t}}{\Delta_R(k)}\hat{u}^L_0(-k\sigma_R/\sigma_L)\ud k\\\nonumber
&+\int_{\partial D^-_0}\frac{-e^{ik(x+2b)-(\sigma_Rk)^2t}(\sigma_L-\sigma_R)-e^{ik(x+2b+2a\sigma_R/\sigma_L)-(\sigma_Rk)^2t}(\sigma_L+\sigma_R)}{2\Delta_R(k)}\hat{u}^R_0(k)\ud k\\\nonumber
&+\int_{\partial D^-_0}\frac{e^{ikx-(\sigma_Rk)^2t}(\sigma_L-\sigma_R)+e^{ik(x+2a\sigma_R/\sigma_L)-(\sigma_Rk)^2t}(\sigma_L+\sigma_R)}{2\Delta_R(k)}\hat{u}^R_0(-k)\ud k\\\nonumber
&+\int_{\partial D^+_0}\frac{2ik\sigma_L\sigma_R^2e^{ik(x+a\sigma_R/\sigma_L)-(\sigma_Rk)^2t}}{\alpha_1\Delta_R(k)}\hat{f}^L((\sigma_R k)^2,t)\ud k\\\nonumber
&+\int_{\partial D^+_0}\frac{-ik\sigma_Re^{ik(x+b)-(\sigma_Rk)^2t}(\sigma_L-\sigma_R)-ik\sigma_R^2e^{ik(x+b+2a\sigma_R/\sigma_L)-(\sigma_Rk)^2t}(\sigma_L+\sigma_R)}{\alpha_3\Delta_R(k)}\hat{f}^R((\sigma_R k)^2,t)\ud k\\\nonumber
&+\int_{\partial D^+_0}\frac{-\sigma_R e^{ikx-(\sigma_Rk)^2t}}{\Delta_R(k)}\hat{u}^L_0(k\sigma_R/\sigma_L)\ud k+
\int_{\partial D^+_0}\frac{\sigma_R e^{ik(x+2a\sigma_R/\sigma_L)-(\sigma_Rk)^2t}}{\Delta_R(k)}\hat{u}^L_0(-k\sigma_R/\sigma_L)\ud k\\\nonumber
&+\int_{\partial D^+_0}\frac{-e^{ikx-(\sigma_Rk)^2t}(\sigma_L+\sigma_R)+e^{ik(x+2a\sigma_R/\sigma_L)-(\sigma_Rk)^2t}(\sigma_R-\sigma_L)}{2\Delta_R(k)}\hat{u}^R_0(k)\ud k\\\label{2isolution}
&+\int_{\partial D^+_0}\frac{e^{ikx-(\sigma_Rk)^2t}(\sigma_L-\sigma_R)+e^{ik(x+2a\sigma_R/\sigma_L)-(\sigma_Rk)^2t}(\sigma_L+\sigma_R)}{2\Delta_R(k)}\hat{u}^R_0(-k)\ud k.
\end{align}

\end{prop}
\vspace*{0.4in}

{\bf Remarks.}

\begin{itemize}

\item The solution of the problem posed in \eqref{doublelayerwall}-\eqref{bcs2} may be obtained using the classical method of separation of variables and superposition, see~\cite{HahnO}. The solutions $u^L(x,t)$ and $u^R(x,t)$ are given by series of eigenfunctions with eigenvalues that satisfy a transcendental equation, closely related to the equation $\Delta_L(k)=0$. This series solution may be obtained from Proposition~\ref{prop:2f} by deforming the contours along $\partial D_0^-$ and $\partial D_0^+$ to the real line, including small semi-circles around each root of either $\Delta_L(k)$ or $\Delta_R(k)$, depending on whether $u^L(x,t)$ or $u^R(x,t)$ is being calculated. Indeed, this is allowed since all integrands decay in the wedges between these contours and the real line, and the zeros of $\Delta_L(k)$ and $\Delta_R(k)$ occur only on the real line, as stated above. Careful calculation of all different contributions, following the examples in~\cite{DeconinckTrogdonVasan, FokasBook}, shows that the contributions along the real line cancel, leaving only residue contributions from the small circles.  Each residue contribution corresponds to a term in the classical series solution. It is not necessarily beneficial to leave the form of the solution in Proposition~\ref{prop:2f} for the series representation, as the latter depends on the roots of $\Delta_L(k)$ and $\Delta_R(k)$, which are not known explicitly. In contrast, the representation of Proposition~\ref{prop:2f} depends on known quantities only and may be readily computed, using one's favorite parameterization of the contours $\partial D_0^-$ and $\partial D_0^+$.

\item Similarly, the familiar piecewise linear steady-state solution of \eqref{doublelayerwall}-\eqref{bcs2} with Dirichlet boundary conditions \cite{HahnO} can be observed from~\eqref{1isolution} and~\eqref{2isolution} by choosing initial conditions that decay appropriately and constant boundary conditions $f_L(t)=\gamma^L$ and $f_R(t)=\gamma^R$. It is convenient to choose zero initial conditions, since the initial conditions do not affect the steady state.  As above, the contours are deformed so that they are along the real line with semi-circular paths around the zeros of $\Delta_L(k)$ and $\Delta_R(k)$, including $k=0$. Since one of these deformations arises from $D_0^+$ while the other comes from $D_0^-$, the contributions along the real line cancel each other, while the semi-circles add to give full residue contributions from the poles associated with the zeros of $\Delta_L(k)$. All such residues vanish as $t\rightarrow \infty$, except at $k=0$. It follows that the steady state behavior is determined by the residue at the origin. This results in

\bea
u^L(x,t)&=&\frac{\sigma_R^2(\gamma^R-\gamma_L)}{b\sigma_L^2+a\sigma_R^2}x+\frac{b\gamma^L\sigma_L^2+
a\gamma^R\sigma_R^2}{b\sigma_L^2+a\sigma_R^2},~~~~~~~~-a<x<0,\\
u^R(x,t)&=&\frac{\sigma_L^2(\gamma^R-\gamma_L)}{b\sigma_L^2+a\sigma_R^2}x+\frac{b\gamma^L\sigma_L^2+
a\gamma^R\sigma_R^2}{b\sigma_L^2+a\sigma_R^2},~~~~~~~~~~~0<x<b,
\eea

\no which is piecewise linear and continuous.

\item A more direct way to recover only the steady-state solution to~\eqref{doublelayerwall} with $\lim_{t\to\infty}f^L(t)=\bar{f}^L$ and $\lim_{t\to\infty}f^R(t)=\bar{f}^R$ constant is to write the solution as the superposition of two parts: $u^L(x,t)=\bar{u}^L(x)+\check{u}^L(x,t)$ and $u^R(x,t)=\bar{u}^R(x)+\check{u}^R(x,t)$. The first parts $\bar u^L$ and $\bar u^R$ satisfy the boundary conditions as $t\rightarrow \infty$ and the stationary heat equation. In other words

\begin{align*}
0&=\sigma_L^2 \bar{u}^L_{xx}(x), && -a<x<0,\\
0&=\sigma_R^2\bar{u}^R_{xx}(x), && 0<x<b,\\
\bar{f}^L&=\alpha_1 \bar{u}^L(-a)+\alpha_2 \bar{u}_x^L(-a), &&\\
\bar{f}^R&=\alpha_3\bar{u}^R(b)+\alpha_4\bar{u}_x^R(b). &&
\end{align*}

\no A piecewise linear ansatz with the imposition of the interface conditions results in linear algebra for the unknown coefficients, see \cite{HahnO}. With the steady state solution in hand, the second (time-dependent) parts $\check u^L$ and $\check u^R$ satisfy the initial conditions modified by the steady state solution and the boundary conditions minus their value as $t\rightarrow \infty$:

\begin{align*}
\check{u}^L_t(x,t)&=\sigma_L^2 \check{u}^L_{xx}(x,t), && a<x<0,~t>0,\\
\check{u}^R_t(x,t)&=\sigma_R^2\check{u}^R_{xx}(x,t), && 0<x<b,~t>0,\\
\check{u}^L(x,0)&=u^L(x,0)-\bar{u}^L(x), && -a<x<0,\\
\check{u}^R(x,0)&=u^R(x,0)-\bar{u}^R(x), && 0<x<b,\\
f^L(t)-\bar{f}^L&=\alpha_1 \check{u}^L(-a,t)+\alpha_2 \check{u}_x^L(-a,t), && t>0,\\
f^R(t)-\bar{f}^R&=\alpha_3\check{u}^R(b,t)+\alpha_4\check{u}_x^R(b,t), && t>0,\\
\end{align*}

where, as usual, we impose continuity of temperature and heat flux at the interface $x=0$. The dynamics of the solution is described by $\check u^L$ and $\check u^R$, both of which decay to zero as $t\rightarrow \infty$. Their explicit form is easily found using the method described in this section.

\end{itemize}

\section{Other problems}\la{sec:other}

With the basic principles of the method outlined in the previous two sections, problems with more layers, both finite and infinite, may be addressed. We state two additional problems below and include solutions for specific initial conditions.  More complete solutions can be found in the electronic supplementary material.

\subsection{Infinite domain with three layers}\label{sec:3i}

In this section we consider the heat equation defined on two semi-infinite rods enclosing a single rod of length~$2a$.

We seek three functions
$$
u^L(x,t),~\;x\in(-\infty,-a),~ \;t\geq0, \qquad u^M(x,t),\;~x\in(-a,b),\;~t\geq 0,\qquad u^R(x,t),\;~x\in(b,\infty),\;~t\geq0,
$$
satisfying the equations
\begin{subequations}\label{3i}
\begin{align}
u^L_t(x,t)&=\sigma_L^2u^L_{xx}(x,t),  &-\infty<&x<-a,&~t>0,\\
{u^M}_t(x,t)&=\sigma^2_M{u^M}_{xx}(x,t),  &-a<&x<a,&~t>0,\\
u^R_t(x,t)&=\sigma_R^2u^R_{xx}(x,t),  &a<&x<\infty,&~t>0,
\end{align}
\end{subequations}

\no the initial conditions
\begin{subequations}\label{3i_ics}
\begin{align}
u^L_t(x,0)&=u^L_{0}(x),  &-\infty<&x<-a,\\
{u^M}_t(x,0)&={u^M}_{0}(x),  &-a<&x<a,\\
u^R_t(x,0)&=u^R_{0}(x),  &a<&x<\infty,
\end{align}
\end{subequations}

\no the asymptotic conditions
\begin{subequations}\label{3i_acs}
\begin{align}
\lim_{x\to-\infty}u^L_t(x,t)&=0,  &t\geq0,\\
\lim_{x\to\infty}u^R_t(x,t)&=0,  &t\geq0,
\end{align}
\end{subequations}

\no and the continuity conditions
\begin{subequations}\label{3i_ccs}
\begin{align}
u^L(-a,t)&=u^M(-a,t),  &t\geq0,\\
u^M(a,t)&=u^R(a,t),  &t\geq0,\\
\sigma_L^2u^L_x(-a,t)&=\sigma_M^2 u^M_x(-a,t),  &t\geq0,\\
\sigma^2_M u^M_x(a,t)&=\sigma_R^2u^R_x(a,t),  &t\geq0,
\end{align}
\end{subequations}

\begin{figure}[tb]
   \centering
   \includegraphics[width=.7\textwidth]{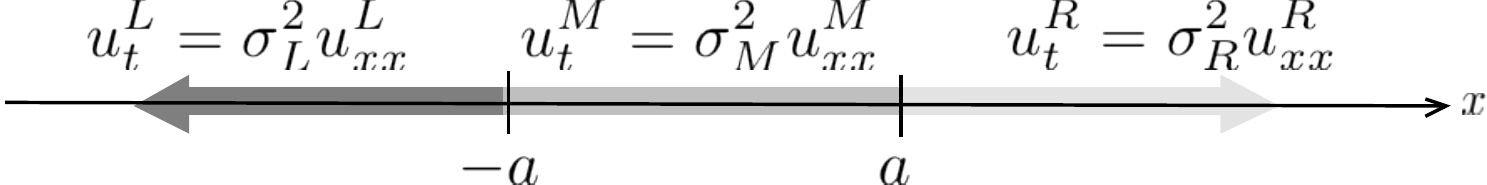} 
   \caption{The heat conduction problem for a single rod of length $2a$ between two semi-infinite rods.    \label{fig:onewall}}
\end{figure}

\no as shown in Figure~\ref{fig:onewall}.

The asymptotic conditions~\eqref{3i_acs} are not the most general possible but are used here to simplify calculations.  To our knowledge, no aspect of this problem is addressed in the literature.

\no To solve this problem, one would proceed with the following steps:

\begin{enumerate}

\item Write the local relations in each of the three domains: left, middle, and right.

\item Use Green's Theorem to define the three global relations, keeping track of where they are valid in the complex $k$ plane.

\item Solve the three global relations for $u^L(x,t)$, $u^M(x,t)$ and $u^R(x,t)$ by inverting the Fourier transforms.

\item Using the $k\to -k$ symmetry of the dispersion relationships on the three original global relations, write down three more global relations. This uses the discrete symmetry of each individual dispersion relation.

\item Deform the integrals of the solution expressions to $D_0^+$ or $D_0^-$ as dictated by the region of analyticity of the integrand.

\item Use the discrete symmetries of the dispersion relationship to the collection of global relations for $g_0(\omega,t)$, $g_1(\omega,t)$, $h_0(\omega,t)$, $h_1(\omega,t)$. Care should be taken that relations can only be solved simultaneously provided their regions of validity overlap. At this stage, as in the previous two sections, the discrete symmetries from one dispersion relation to another come into play.

\item Terms containing unknown functions are shown to be zero by examining the regions of analyticity and decay for the relevant integrands, and the use of Jordan's Lemma.

\end{enumerate}

For brevity of the presentation, we will assume $u^M(x,0)=0=u^R(x,0)$.  The solution to the non-restricted problem can be found in the electronic supplementary material.  After defining the transforms

\begin{align*}
&&\hat{u}^L_0(k)&=\int_{-\infty}^{-a}e^{-ikx}u^L_0(x)\ud x,
&\hat{u}^L(k,t)&=\int_{-\infty}^{-a}e^{-ikx}u^L(x,t)\ud x,\\
&&\hat{u}^M(k)&=\int_{-a}^{a} e^{-ikx}{u^M}(x,t)\ud x,
&\hat{u}^R(k,t)&=\int_{a}^{\infty} e^{-ikx}u^R(x,t)\ud x,\\
&&h_1(\omega,t)&=\int_{0}^te^{\omega s}u^L_x(-a,s)\ud s=\frac{\sigma_M^2}{\sigma_L^2}\int_{0}^te^{\omega s}u^M_x(-a,s)\ud s, \\
&&h_0(\omega,t)&=\int_{0}^te^{\omega s}u^L(-a,s)\ud s=\int_{0}^te^{\omega s}u^M(-a,s)\ud s,\\
&&g_{1}(\omega,t)&=\int_{0}^te^{\omega s}u^M_x(a,s)\ud s=\frac{\sigma_R^2}{\sigma_M^2}\int_{0}^te^{\omega s}u^R_x(a,s)\ud s,\\
&&g_{0}(\omega,t)&=\int_{0}^te^{\omega s}u^M(a,s)\ud s=\int_{0}^te^{\omega s}u^R(a,s)\ud s,
\end{align*}

we proceed as outlined above. The solution formulae are given in the following proposition:

\begin{prop}
The solution of the heat transfer problem~\eqref{3i}-\eqref{3i_ccs} with $u^M(x,0)=u^R(x,0)=0$ is

\begin{align*}
u^L(x,t)&=\frac{1}{2\pi}\int_{-\infty}^\infty  e^{ikx-(\sigma_Lk)^2t}\hat{u}_0^L(k)\ud k\\
&-\frac{1}{2}\int_{\partial D_0^-}\frac{e^{ik(x+2a)-(\sigma_L k)^2t}}{\Delta_L(k)}\left(  (\sigma_L+\sigma_M)(\sigma_M-\sigma_R)+e^{4ika\sigma_L/\sigma_M}(\sigma_L-\sigma_M)(\sigma_M+\sigma_R)\right)\hat{u}_0^L(-k)\ud k,
\end{align*}

\no for $-\infty<x<-a$ with $\Delta_L(k)=\pi\left((\sigma_L-\sigma_M)(\sigma_M-\sigma_R)+e^{4iak\sigma_L/\sigma_M}(\sigma_L+\sigma_M)(\sigma_M+\sigma_R)\right)$,

\begin{align*}
u^M(x,t)&=-\sigma_M(\sigma_M-\sigma_R)\int_{\partial D_0^-}\frac{e^{ik(x+a+a\sigma_M/\sigma_L)-(\sigma_M k)^2t}}{\Delta_L(k\sigma_M/\sigma_L)}\hat{u}_0^L\left(\frac{-k\sigma_M}{\sigma_L}\right)\ud k\\
&+\sigma_M(\sigma_M+\sigma_R)\!\!\int_{\partial D_0^+}\frac{e^{ik(x+a-a\sigma_M/\sigma_L)-(\sigma_Mk)^2t}}{\Delta_R(k\sigma_M/\sigma_R)}\hat{u}^L_0\left(\frac{k\sigma_M}{\sigma_L}\right)\ud k,
\end{align*}

\no for $-a<x<b$ with $\Delta_R(k)=\pi\left((\sigma_L+\sigma_M)(\sigma_M+\sigma_R)+e^{4iak\sigma_R/\sigma_M}(\sigma_L-\sigma_M)(\sigma_M-\sigma_R)\right)$. Lastly, the expression for $u^R(x,t)$, valid for $x>a$, is

$$
u^R(x,t)=2\sigma_M\sigma_R\int_{\partial D_0^+}\frac{e^{ik(x-a-a\sigma_R/\sigma_L+2a\sigma_R/\sigma_M)-(\sigma_Rk)^2t}}{\Delta_R(k)} \hat{u}^L_0\left(\frac{k\sigma_R}{\sigma_L}\right)\ud k.
$$
\end{prop}

\subsection{Finite domain with three layers}

\begin{figure}[htbp]
   \centering
   \includegraphics[width=.7\textwidth]{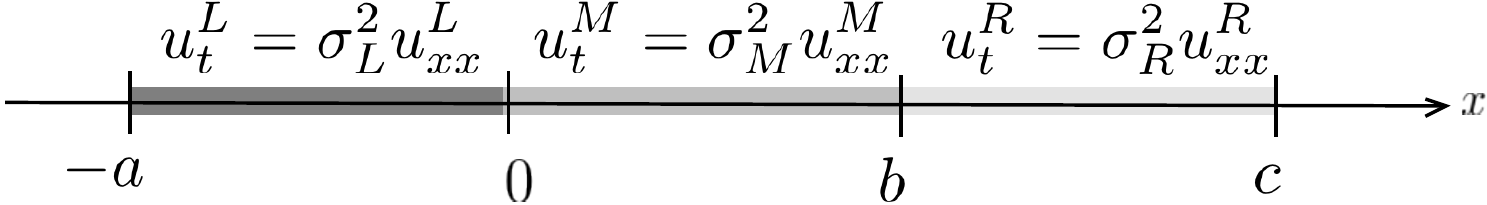} 
   \caption{The heat equation for three finite layers.    \label{fig:threewall}}
\end{figure}

We consider the heat conduction problem in three rods of finite length as shown in Figure~\ref{fig:threewall}.

We seek three functions
$$
u^L(x,t),~\;x\in(-a,0),~ \;t\geq0, \qquad u^M(x,t),\;~~x\in(0,b),\;~~t\geq 0,\qquad u^R(x,t),\;~~x\in(b,c),\;~~t\geq0,
$$
satisfying the equations
\begin{subequations}\label{3f}
\begin{align}
u^L_t(x,t)&=\sigma_L^2u^L_{xx}(x,t),  &-a<&x<0,&~t>0,\\
{u^M}_t(x,t)&=\sigma^2_M{u^M}_{xx}(x,t),  &0<&x<b,&~t>0,\\
u^R_t(x,t)&=\sigma_R^2u^R_{xx}(x,t),  &b<&x<c,&~t>0,
\end{align}
\end{subequations}

\no the initial conditions
\begin{subequations}\label{3f_ics}
\begin{align}
u^L_t(x,0)&=u^L_{0}(x),  &-\infty<&x<-a,\\
{u^M}_t(x,0)&={u^M}_{0}(x),  &-a<&x<a,\\
u^R_t(x,0)&=u^R_{0}(x),  &a<&x<\infty,
\end{align}
\end{subequations}

\no the boundary conditions
\begin{subequations}\label{3f_bcs}
\begin{align}
f^L(t)&=\alpha_1 u^L(-a,t)+\alpha_2 u^L_x(-a,t), && &t>0,\\
f^R(t)&=\alpha_3 u^R(c,t)+\alpha_4 u^R_x(c,t), && &t>0,\\
\end{align}
\end{subequations}

\no and the continuity conditions
\begin{subequations}\label{3f_ccs}
\begin{align}
u^L(0,t)&=u^M(0,t),  &t\geq0,\\
u^M(b,t)&=u^R(b,t),  &t\geq0,\\
\sigma_L^2u^L_x(0,t)&=\sigma_M^2 u^M_x(0,t),  &t\geq0,\\
\sigma^2_M u^M_x(b,t)&=\sigma_R^2u^R_x(b,t),  &t\geq0.
\end{align}
\end{subequations}

The solution process is as before, following the steps outlined in the previous section.  For simplicity we assume Neumann boundary data ($\alpha_1=\alpha_3=0$), zero boundary conditions ($f^L(t)=f^R(t)=0$), and $u^M(x,0)=0=u^R(x,0)$.  The solution with $u^M(x,0)\neq0$ and $u^R(x,0)\neq0$ is given in the electronic supplementary material. We define

\begin{align*}
&&\hat{u}^L_0(k)&=\int_{-a}^{0}e^{-ikx}u^L_0(x)\ud x,
&\hat{u}^L(k,t)&=\int_{-a}^{0}e^{-ikx}u^L(x,t)\ud x,\\
&&\hat{u}^M(k)&=\int_{0}^{b} e^{-ikx}{u^M}(x,t)\ud x,
&\hat{u}^R(k,t)&=\int_{b}^{c} e^{-ikx}u^R(x,t)\ud x,\\
&&g^L_{1}(\omega,t)&=\int_{0}^te^{\omega s}u^L_x(-a,s)\ud s,
&g^L_{0}(\omega,t)&=\int_{0}^te^{\omega s}u^L(-a,s)\ud s,\\
&&h^L_1(\omega,t)&=\int_{0}^te^{\omega s}u^L_x(0,s)\ud s=\frac{\sigma_M^2}{\sigma_L^2}\int_{0}^te^{\omega s}u^M_x(0,s)\ud s,\\
&&h^L_0(\omega,t)&=\int_{0}^te^{\omega s}u^L(0,s)\ud s=\int_{0}^te^{\omega s}u^M(0,s)\ud s,\\
&&g^R_{1}(\omega,t)&=\int_{0}^te^{\omega s}u^M_x(b,s)\ud s=\frac{\sigma_R^2}{\sigma_M^2}\int_{0}^te^{\omega s}u^R_x(b,s)\ud s,\\
&&g^R_{0}(\omega,t)&=\int_{0}^te^{\omega s}u^M(b,s)\ud s=\int_{0}^te^{\omega s}u^R(b,s)\ud s,\\
&&h^R_1(\omega,t)&=\int_{0}^te^{\omega s}u^R_x(c,s)\ud s,
&h^R_0(\omega,t)&=\int_{0}^te^{\omega s}u^R(c,s)\ud s.\\
\end{align*}

The solution is given by the following proposition.

\begin{prop}
The solution to the heat transfer problem~\eqref{3f}-\eqref{3f_ccs} with $\alpha_1=\alpha_3=0$, $f^L(t)=f^R(t)=0$, and $u^M(x,0)=u^R(x,0)=0$ is

\begin{align*}
&u^L(x,t)=\frac{1}{2\pi}\int_{-\infty}^\infty e^{i k x-(\sigma_L k)^2t}\hat{u}_0^L(k)\ud k\\
&+\int_{\partial D_0^-}\!\!\!\!\!\! \frac{e^{ikx-(\sigma_Lk)^2t}}{2\Delta_L(k)} \left(e^{2ibk\sigma_L/\sigma_R}(\sigma_L\!-\!\sigma_M)(\sigma_M\!-\!\sigma_R)+e^{2ik\sigma_L(c/\sigma_R+
b/\sigma_M)}(\sigma_L\!+\!\sigma_M)(\sigma_M\!-\!\sigma_R) \right.\\
&~~+\left.e^{2ick\sigma_L/\sigma_R}(\sigma_L-\sigma_M)(\sigma_M+\sigma_R)+e^{2ibk\sigma_L(1/\sigma_R+1/\sigma_M)}(\sigma_L+\sigma_M)(\sigma_M+\sigma_R)   \right)  \hat{u}_0^L(k) \ud k\\
&+\int_{\partial D_0^-}\!\!\!\!\!\!\!\!  \frac{e^{ik(x+2a)-(\sigma_Lk)^2t}}{2\Delta_L(k)} \!\left(\! e^{2ibk\sigma_L/\sigma_R}(\sigma_L\!-\!\sigma_M)(\sigma_M\!-\!\sigma_R)\!+\!e^{2ik\sigma_L(c/\sigma_R+
b/\sigma_M)}(\sigma_L\!+\!\sigma_M)(\sigma_M\!-\!\sigma_R)+\!\right.\\
&~~+\left.e^{2ick\sigma_L/\sigma_R}(\sigma_L-\sigma_M)(\sigma_M+\sigma_R)+e^{2ibk\sigma_L(1/\sigma_R+1/\sigma_M)}(\sigma_L+\sigma_M)(\sigma_M+\sigma_R)   \right)  \hat{u}_0^L(-k) \ud k\\
&+\int_{\partial D_0^+}\!\!\!\!\!\!\!\! \frac{e^{ik(x+2a)\!-\!(\sigma_Lk)^2t}}{2\Delta_L(k)}\!\left(\!e^{2ik\sigma_L(c/\sigma_R\!+\!b/\sigma_M)} (\sigma_L\!-\!\sigma_M)(\sigma_M\!-\!\sigma_R)\!+\!e^{2ibk\sigma_L/\sigma_R}(\sigma_L\!+\!\sigma_M)(\sigma_M\!-\!\sigma_R)\!+\!\right.\\
&~~+\left.e^{2ibk\sigma_L(1/\sigma_R+1/\sigma_M)}(\sigma_L-\sigma_M)(\sigma_M+\sigma_R)+e^{2ick\sigma_L/\sigma_R}(\sigma_L+\sigma_M)(\sigma_M+\sigma_R)    \right)  \hat{u}_0^L(k)\ud k \\
&+\int_{\partial D_0^+} \!\!\!\!\!\!\!\! \frac{e^{ik(x+2a)\!-\!(\sigma_Lk)^2t}}{2\Delta_L(k)} \left(e^{2ibk\sigma_L/\sigma_R}(\sigma_L\!-\!\sigma_M)(\sigma_M\!-\!\sigma_R)\!+\!e^{2ik\sigma_L(c/\sigma_R\!+\!b/\sigma_M)}(\sigma_L\!+\!\sigma_M)(\sigma_M\!-\!\sigma_R) \right.\\
&~~+\left.e^{2ick\sigma_L/\sigma_R}(\sigma_L-\sigma_M)(\sigma_M+\sigma_R)+e^{2ibk\sigma_L(1/\sigma_R+1/\sigma_M)}(\sigma_L+\sigma_M)(\sigma_M+\sigma_R)       \right) \hat{u}_0^L(-k)\ud k,
\end{align*}

\no for $-a<x<0$ with

\bea
\Delta_L(k)&=&\pi \left(e^{2ibk\sigma_L/\sigma_R}(\sigma_L-\sigma_M)(\sigma_M-\sigma_R)+
e^{2ik(c\sigma_L/\sigma_R+b\sigma_L/\sigma_M+a)}(\sigma_M-\sigma_L)(\sigma_M-\sigma_R) \right.\\
&&+e^{2ik\sigma_L(c/\sigma_R+b/\sigma_M)}(\sigma_L+\sigma_M)(\sigma_M-\sigma_R)+
e^{2ik(b\sigma_L/\sigma_R+a)}(\sigma_L+\sigma_M)(\sigma_R-\sigma_M)\\
&&+e^{2ick\sigma_L/\sigma_R}(\sigma_L-\sigma_M)(\sigma_M+\sigma_R)+
e^{2ik(a+b\sigma_L/\sigma_R+b\sigma_L/\sigma_M)}(\sigma_M-\sigma_L)(\sigma_M+\sigma_R)\\
&&\left.+e^{2ibk(\sigma_L/\sigma_R+\sigma_L/\sigma_M)}(\sigma_L+\sigma_M)(\sigma_M+
\sigma_R)-e^{2ik(c\sigma_L/\sigma_R+a)}(\sigma_L+\sigma_M)(\sigma_M+\sigma_R)\right).
\eea

\no Next,

\begin{align*}
&u^M(x,t)=\!\int_{\partial D_0^-}   \frac{-e^{ik(x+b)-(\sigma_Mk)^2t}\sigma_M}{\Delta_M(k)} \left(e^{2ick\sigma_M/\sigma_R}(\sigma_M-\sigma_R)+e^{2ibk\sigma_M/\sigma_R}(\sigma_M+\sigma_R)   \right)  \hat{u}_0^L(k\sigma_M/\sigma_L) \ud k\\
&\!+\!\int_{\partial D_0^-} \!\!\!\!\!\! \frac{-e^{ik(x+b+2a\sigma_M/\sigma_L)-(\sigma_Mk)^2t}\sigma_M}{\Delta_M(k)} \!\left(\!e^{2ick\sigma_M/\sigma_R}(\sigma_M\!-\!\sigma_R)\!+\!e^{2ibk\sigma_M/\sigma_R}(\sigma_M\!+\!\sigma_R) \!\right)\!  \hat{u}_0^L(-\sigma_M/\sigma_L)\! \ud k\\
&+\int_{\partial D_0^+} \frac{-e^{ik(x+b)-(\sigma_Mk)^2t}\sigma_M}{\Delta_M(k)}\left(e^{2ick\sigma_M/\sigma_R}(\sigma_M-\sigma_R)+e^{2ibk\sigma_M/\sigma_R}(\sigma_M+\sigma_R)  \right)  \hat{u}_0^L(k\sigma_M/\sigma_L)\ud k\\
&\!-\!\int_{\partial D_0^+}\!\!\!\!\!\!  \frac{e^{ik(x+b+2a\sigma_M/\sigma_L)-(\sigma_Mk)^2t}\sigma_M}{\Delta_M(k)}\!\left(\!e^{2ick\sigma_M/\sigma_R}(\sigma_M\!-\!\sigma_R)\!+\!e^{2ibk\sigma_M/\sigma_R}(\sigma_M\!+\!\sigma_R)  \!\right)\! \hat{u}_0^L(-k\sigma_M/\sigma_L)\!\ud k,
\end{align*}

\no for $0<x<b$, with

\bea
\Delta_M(k)&=&\pi \left(e^{ik(a\sigma_M/\sigma_L+b+c\sigma_M/\sigma_R)}(\sigma_L-\sigma_M)(\sigma_M-\sigma_R)+e^{2bik\sigma_M/\sigma_R}(\sigma_M-\sigma_L)(\sigma_M-\sigma_R) \right.\\
&&+e^{2ik\sigma_M(b/\sigma_R+a/\sigma_L)}(\sigma_L+\sigma_M)(\sigma_M-\sigma_R)+e^{2ik(c\sigma_M/\sigma_R+b)}(\sigma_L+\sigma_M)(\sigma_R-\sigma_M)\\
&&+e^{2ik(a\sigma_M/\sigma_L+b\sigma_M/\sigma_R+b)}(\sigma_L-\sigma_M)(\sigma_M+\sigma_R)+e^{2ick\sigma_M/\sigma_R}(\sigma_M-\sigma_L)(\sigma_M+\sigma_R)\\
&&\left.-e^{2ibk(\sigma_M/\sigma_R+1)}(\sigma_L+\sigma_M)(\sigma_M+\sigma_R)+e^{2ik\sigma_M(c/\sigma_R+a/\sigma_L)}(\sigma_L+\sigma_M)(\sigma_M+\sigma_R)\right),
\eea
and

\begin{align*}
&u^R(x,t)=\int_{\partial D_0^-}   \frac{-2e^{ik(x+b+b\sigma_R/\sigma_M)-(\sigma_Rk)^2t}}{\Delta_R(k)} \left(\sigma_M\sigma_R   \right)  \hat{u}_0^L(k\sigma_R/\sigma_L) \ud k\\
&+\int_{\partial D_0^-}  \frac{-2e^{ik(x+b+b\sigma_R/\sigma_M+2a\sigma_R/\sigma_L)-(\sigma_Rk)^2t}}{\Delta_R(k)} \left(\sigma_M\sigma_R      \right)  \hat{u}_0^L(-k\sigma_R/\sigma_L) \ud k\\
&+\int_{\partial D_0^+} \frac{2e^{ik(x+b+b\sigma_R/\sigma_M)-(\sigma_Rk)^2t}}{\Delta_R(k)}\left(\sigma_M\sigma_R     \right)  \hat{u}_0^L(k\sigma_R/\sigma_L)\ud k \\
&+\int_{\partial D_0^+}  \frac{2e^{ik(x+b+b\sigma_R/\sigma_M+2a\sigma_R/\sigma_L)-(\sigma_Rk)^2t}}{\Delta_R(k)} \left(  \sigma_M\sigma_R    \right) \hat{u}_0^L(-k\sigma_R/\sigma_L)\ud k,
\end{align*}

\no for $b<x<c$ with

\begin{align*}
\Delta_R(k)&=\pi \left(e^{2ibk}(\sigma_L-\sigma_M)(\sigma_M-\sigma_R)+e^{2ik(a\sigma_R/\sigma_L+b\sigma_R/\sigma_M+c)}(\sigma_M-\sigma_L)(\sigma_M-\sigma_R) \right.\\
&+e^{2ik(b\sigma_R/\sigma_M+c)}(\sigma_L+\sigma_M)(\sigma_M-\sigma_R)+e^{2ik(a\sigma_R/\sigma_L+b)}(\sigma_L+\sigma_M)(\sigma_R-\sigma_M)\\
&+e^{2ick}(\sigma_L-\sigma_M)(\sigma_M+\sigma_R)+e^{2ik(a/\sigma_L+b+b\sigma_R/\sigma_M)}(\sigma_M-\sigma_L)(\sigma_M+\sigma_R)\\
&\left.+e^{2ibk(1+\sigma_R/\sigma_M)}(\sigma_L+\sigma_M)(\sigma_M+\sigma_R)-e^{2ik(a\sigma_R/\sigma_L+c)}(\sigma_L+\sigma_M)(\sigma_M+\sigma_R)\right).
\end{align*}
\end{prop}

\section*{Acknowledgements}

This work was generously supported by the National Science Foundation under grant NSF-DMS-1008001 (BD, NES).  Natalie Sheils also acknowledges support from the National Science Foundation under grant number NSF-DGE-0718124.  Any opinions, findings, and conclusions or recommendations expressed in this material are those of the authors and do not necessarily reflect the views of the funding sources.

We wish to thank the American Institute for Mathematics (AIM, Palo Alto, CA) for its hospitality, while some of this work was being conducted.

\newpage
\appendix
\section{Electronic Supplementary Material}

\subsection{Infinite domain with three layers}\label{Asec:3i}
In this section we consider the heat equation defined on two semi-infinite rods enclosing a single rod of length~$2a$ as defined in the main paper,~\eqref{3i}-\eqref{3i_ccs} without imposing $u^M(x,0)=0=u^R(x,0)$.  That is, we seek three functions
$$
u^L(x,t),~\;x\in(-\infty,-a),~ \;t\geq0, \qquad u^M(x,t),\;~~x\in(-a,b),\;~~t\geq 0,\qquad u^R(x,t),\;~~x\in(b,\infty),\;~~t\geq0,
$$
satisfying the equations
\begin{subequations}\label{A3i}
\begin{align}
u^L_t(x,t)&=\sigma_L^2u^L_{xx}(x,t),  &-\infty<&x<-a,&~t>0,\\
{u^M}_t(x,t)&=\sigma^2_M{u^M}_{xx}(x,t),  &-a<&x<a,&~t>0,\\
u^R_t(x,t)&=\sigma_R^2u^R_{xx}(x,t),  &a<&x<\infty,&~t>0,
\end{align}
\end{subequations}

the initial conditions
\begin{subequations}\label{A3i_ics}
\begin{align}
u^L_t(x,0)&=u^L_{0}(x),  &-\infty<&x<-a,\\
{u^M}_t(x,0)&={u^M}_{0}(x),  &-a<&x<a,\\
u^R_t(x,0)&=u^R_{0}(x),  &a<&x<\infty,
\end{align}
\end{subequations}

the asymptotic conditions
\begin{subequations}\label{A3i_acs}
\begin{align}
\lim_{x\to-\infty}u^L_t(x,t)&=0,  &t\geq0,\\
\lim_{x\to\infty}u^R_t(x,t)&=0,  &t\geq0,
\end{align}
\end{subequations}

and the continuity conditions
\begin{subequations}\label{A3i_ccs}
\begin{align}
&u^L(-a,t)=u^M(-a,t),  &t\geq0,\\
&u^M(a,t)=u^R(a,t),  &t\geq0,\\
&\sigma_L^2u^L_x(-a,t)=\sigma_M^2 u^M_x(-a,t),  &t\geq0,\\
&\sigma^2_M u^M_x(a,t)=\sigma_R^2u^R_x(a,t),  &t\geq0.
\end{align}
\end{subequations}

After defining the transforms

\begin{align*}
&&\hat{u}^L_0(k)&=\int_{-\infty}^{-a}e^{-ikx}u^L_0(x)\ud x,
&\hat{u}^L(k,t)&=\int_{-\infty}^{-a}e^{-ikx}u^L(x,t)\ud x,\\
&&\hat{u}^M_0(k)&=\int_{-a}^{a} e^{-ikx}u^M_0(x)\ud x,
&\hat{u}^M(k)&=\int_{-a}^{a} e^{-ikx}{u^M}(x,t)\ud x,\\
&&\hat{u}^R_0(k)&=\int_{a}^{\infty} e^{-ikx}u^R_0(x)\ud x,
&\hat{u}^R(k,t)&=\int_{a}^{\infty} e^{-ikx}u^R(x,t)\ud x,\\
&&h_1(\omega,t)&=\int_{0}^te^{\omega s}u^L_x(-a,s)\ud s=\frac{\sigma_M^2}{\sigma_L^2}\int_{0}^te^{\omega s}u^M_x(-a,s)\ud s, \\
&&h_0(\omega,t)&=\int_{0}^te^{\omega s}u^L(-a,s)\ud s=\int_{0}^te^{\omega s}u^M(-a,s)\ud s,\\
&&g_{1}(\omega,t)&=\int_{0}^te^{\omega s}u^M_x(a,s)\ud s=\frac{\sigma_R^2}{\sigma_M^2}\int_{0}^te^{\omega s}u^R_x(a,s)\ud s,\\
&&g_{0}(\omega,t)&=\int_{0}^te^{\omega s}u^M(a,s)\ud s=\int_{0}^te^{\omega s}u^R(a,s)\ud s,
\end{align*}

we proceed as outlined in the preceding sections. The solution formulae are given in the following proposition:

\begin{prop}
The solution of the heat transfer problem~\eqref{A3i}-\eqref{A3i_ccs} is

\begin{align*}
&u^L\!(x,t)\!=\!\frac{1}{2\pi}\!\!\int_{-\infty}^\infty \!\!\!\! e^{ikx-(\sigma_Lk)^2t}\hat{u}_0^L(k)\ud k
\!-\!\sigma_L(\sigma_M+\sigma_R)\!\!\int_{\partial D_0^-}\!\!\!\!\!\!\!\!\frac{e^{ik(x+a+3a\sigma_L/\sigma_M)-(\sigma_L k)^2t}}{\Delta_L(k)}\hat{u}_0^M\!\left(\!\frac{k\sigma_L}{\sigma_M}\!\right)\!\ud k\\
&-\frac{1}{2}\int_{\partial D_0^-}\!\!\!\!\!\!\!\!\frac{e^{ik(x+2a)-(\sigma_L k)^2t}}{\Delta_L(k)}\left(  (\sigma_L+\sigma_M)(\sigma_M-\sigma_R)+e^{4ika\sigma_L/\sigma_M}(\sigma_L-\sigma_M)(\sigma_M+\sigma_R)\right)\hat{u}_0^L(-k)\ud k\\
&\sigma_L(\sigma_R\!-\!\sigma_M)\!\!\int_{\partial D_0^-}\!\!\!\!\!\!\!\!\!\!\frac{e^{ik(x+a+a\sigma_L/\sigma_M)\!-\!(\sigma_L k)^2t}}{\Delta_L(k)}\hat{u}_0^M\!\!\left(\!\frac{-k\sigma_L}{\sigma_M}\!\!\right)\!\!\!\ud k
\!-\!2\sigma_L\sigma_M\!\!\!\int_{\partial D_0^-}\!\!\!\!\!\!\!\!\!\!\frac{e^{ik(x+a+a\sigma_L/\sigma_R+2a\sigma_L/\sigma_M)\!-\!(\sigma_L k)^2t}}{\Delta_L(k)}\hat{u}_0^R\!\!\left(\!\!\frac{k\sigma_L}{\sigma_R}\!\!\right)\!\!\!\ud k,
\end{align*}

\no for $-\infty<x<-a$ with $\Delta_L(k)=\pi\left((\sigma_L-\sigma_M)(\sigma_M-\sigma_R)+e^{4iak\sigma_L/\sigma_M}(\sigma_L+\sigma_M)(\sigma_M+\sigma_R)\right)$,

\begin{align*}
&u^M\!(x,t)\!=\!\frac{1}{2}\!\!\int_{-\infty}^\infty \!\!\!\! e^{ikx-(\sigma_Mk)^2t} \hat{u}_0^M(k)
\!-\!\sigma_M(\sigma_M\!-\!\sigma_R)\!\!\int_{\partial D_0^-}\!\!\!\!\frac{e^{ik(x+a+a\sigma_M/\sigma_L)-(\sigma_M k)^2t}}{\Delta_L(k\sigma_M/\sigma_L)}\hat{u}_0^L\!
\left(\!\frac{-k\sigma_M}{\sigma_L}\!\right)\!\ud k\\
&+\frac{(\sigma_L\!-\!\sigma_M)(\sigma_M\!-\!\sigma_R)}{2}\!\!\int_{\partial D_0^-}\!\!\!\!\!\!\frac{e^{ikx-(\sigma_M k)^2t}\hat{u}_0^M\!(k)\!}{\Delta_L(k\sigma_M/\sigma_L)}\ud k
\!+\!\sigma_M(\sigma_M\!+\!\sigma_R)\!\!\int_{\partial D_0^+}\!\!\!\!\!\!\!\!\frac{e^{ik(x+a-a\sigma_M/\sigma_L)\!-\!(\sigma_Mk)^2t}}{\Delta_R(k\sigma_M/\sigma_R)}\hat{u}^L_0\!\!\left(\!\frac{k\sigma_M}{\sigma_L}\!\right)\!\!\ud k\\
&-\sigma_M(\sigma_L\!+\!\sigma_M)\!\!\int_{\partial D_0^-}\!\!\!\!\!\!\!\!\frac{e^{ik(x+3a+a\sigma_M/\sigma_R)\!-\!(\sigma_M k)^2t}}{\Delta_L(k\sigma_M/\sigma_L)}
\hat{u}_0^R\!\!\left(\frac{k\sigma_M}{\sigma_R}\right)\!\!\ud k\!
-\!\frac{(\sigma_L\!+\!\sigma_M)(\sigma_M\!-\!\sigma_R)}{2}\!\!\int_{\partial D_0^-}\!\!\!\!\!\!\!\!\frac{e^{ik(x+2a)-(\sigma_M k)^2t}\hat{u}_0^M\!(-k)}{\Delta_L(k\sigma_M/\sigma_L)}\ud k\\
&+\frac{(\sigma_M\!-\!\sigma_L)(\sigma_M\!+\!\sigma_R)}{2}\!\!\int_{\partial D_0^+}\!\!\!\!\!\!\!\!\frac{e^{ik(x+2a)-(\sigma_Mk)^2t}\hat{u}^M_0(k)}{\Delta_R(k\sigma_M/\sigma_R)}\ud k
\!+\!\frac{(\sigma_M\!-\!\sigma_L)(\sigma_M\!-\!\sigma_R)}{2}\!\!\int_{\partial D_0^+}\!\!\!\!\!\!\!\!\frac{e^{ik(x+4a)-(\sigma_Mk)^2t}\hat{u}^M_0(-k)}{\Delta_R(k\sigma_M/\sigma_R)}\ud k\\
&+\sigma_M(\sigma_M-\sigma_L)\int_{\partial D_0^+}\frac{e^{ik(x+3a-a\sigma_M/\sigma_R)-
(\sigma_Mk)^2t}}{\Delta_R(k\sigma_M/\sigma_R)}
\hat{u}^R_0\!\left(\!\frac{-k\sigma_M}{\sigma_R}\!\right)\!\ud k,\end{align*}

\no for $-a<x<b$ with $\Delta_R(k)=\pi\left((\sigma_L+\sigma_M)(\sigma_M+\sigma_R)+e^{4iak\sigma_R/\sigma_M}(\sigma_L-
\sigma_M)(\sigma_M-\sigma_R)\right)$. Lastly, the expression for $u^R(x,t)$, valid for $x>a$, is identical to that for $u^L(x,t)$ with the replacements $a\leftrightarrow -a$, $R\leftrightarrow L$, and $\partial D_0^- \leftrightarrow -\partial D_0^+$.

%

\end{prop}

\subsection{Finite domain with three layers}

In this section we consider the heat conduction problem in three rods of finite length as defined in the main paper by~\eqref{3f}-\eqref{3f_ccs} without assuming $u^M(x,0)=0=u^R(x,0)$.  That is, we seek three functions
$$
u^L(x,t),~\;x\in(-a,0),~ \;t\geq0, \qquad u^M(x,t),\;~~x\in(0,b),\;~~t\geq 0,\qquad u^R(x,t),\;~~x\in(b,c),\;~~t\geq0,
$$
satisfying the equations
\begin{subequations}\label{A3f}
\begin{align}
u^L_t(x,t)&=\sigma_L^2u^L_{xx}(x,t),  &-a<&x<0,&~t>0,\\
{u^M}_t(x,t)&=\sigma^2_M{u^M}_{xx}(x,t),  &0<&x<b,&~t>0,\\
u^R_t(x,t)&=\sigma_R^2u^R_{xx}(x,t),  &b<&x<c,&~t>0,
\end{align}
\end{subequations}

the initial conditions
\begin{subequations}\label{A3f_ics}
\begin{align}
u^L_t(x,0)&=u^L_{0}(x),  &-\infty<&x<-a,\\
{u^M}_t(x,0)&={u^M}_{0}(x),  &-a<&x<a,\\
u^R_t(x,0)&=u^R_{0}(x),  &a<&x<\infty,
\end{align}
\end{subequations}

the boundary conditions
\begin{subequations}\label{A3f_bcs}
\begin{align}
f^L(t)&=\alpha_1 u^L(-a,t)+\alpha_2 u^L_x(-a,t), && &t>0,\\
f^R(t)&=\alpha_3 u^R(c,t)+\alpha_4 u^R_x(c,t), && &t>0,\\
\end{align}
\end{subequations}

and the continuity conditions
\begin{subequations}\label{A3f_ccs}
\begin{align}
&u^L(0,t)=u^M(0,t),  &t\geq0,\\
&u^M(b,t)=u^R(b,t),  &t\geq0,\\
&\sigma_L^2u^L_x(0,t)=\sigma_M^2 u^M_x(0,t),  &t\geq0,\\
&\sigma^2_M u^M_x(b,t)=\sigma_R^2u^R_x(b,t),  &t\geq0.
\end{align}
\end{subequations}

The solution process is as before, following the steps outlined in Section~{sec:3i} of the main paper. For simplicity we assume Neumann boundary data ($\alpha_1=\alpha_3=0$) and zero boundary conditions ($f^L(t)=f^R(t)=0$). We define

\begin{align*}
&&\hat{u}^L_0(k)&=\int_{-a}^{0}e^{-ikx}u^L_0(x)\ud x,
&\hat{u}^L(k,t)&=\int_{-a}^{0}e^{-ikx}u^L(x,t)\ud x,\\
&&\hat{u}^M_0(k)&=\int_{0}^{b} e^{-ikx}u^M_0(x)\ud x,
&\hat{u}^M(k)&=\int_{0}^{b} e^{-ikx}{u^M}(x,t)\ud x,\\
&&\hat{u}^R_0(k)&=\int_{b}^{c} e^{-ikx}u^R_0(x)\ud x,
&\hat{u}^R(k,t)&=\int_{b}^{c} e^{-ikx}u^R(x,t)\ud x,\\
&&g^L_{1}(\omega,t)&=\int_{0}^te^{\omega s}u^L_x(-a,s)\ud s,
&g^L_{0}(\omega,t)&=\int_{0}^te^{\omega s}u^L(-a,s)\ud s,\\
&&h^L_1(\omega,t)&=\int_{0}^te^{\omega s}u^L_x(0,s)\ud s=\frac{\sigma_M^2}{\sigma_L^2}\int_{0}^te^{\omega s}u^M_x(0,s)\ud s,\\
&&h^L_0(\omega,t)&=\int_{0}^te^{\omega s}u^L(0,s)\ud s=\int_{0}^te^{\omega s}u^M(0,s)\ud s,\\
&&g^R_{1}(\omega,t)&=\int_{0}^te^{\omega s}u^M_x(b,s)\ud s=\frac{\sigma_R^2}{\sigma_M^2}\int_{0}^te^{\omega s}u^R_x(b,s)\ud s,\\
&&g^R_{0}(\omega,t)&=\int_{0}^te^{\omega s}u^M(b,s)\ud s=\int_{0}^te^{\omega s}u^R(b,s)\ud s,\\
&&h^R_1(\omega,t)&=\int_{0}^te^{\omega s}u^R_x(c,s)\ud s,
&h^R_0(\omega,t)&=\int_{0}^te^{\omega s}u^R(c,s)\ud s.\\
\end{align*}

The solution is given by the following proposition.

\begin{prop}
The solution to the heat transfer problem~\eqref{A3f}-\eqref{A3f_ccs} with $\alpha_1=\alpha_3=0$ and $f^L(t)=f^R(t)=0$ is

\begin{align*}
&u^L(x,t)=\frac{1}{2\pi}\int_{-\infty}^\infty e^{i k x-(\sigma_L k)^2t}\hat{u}_0^L(k)\ud k\\
&+\int_{\partial D_0^-}\!\!\!\!\!\! \frac{e^{ikx-(\sigma_Lk)^2t}}{2\Delta_L(k)} \left(e^{2ibk\sigma_L/\sigma_R}(\sigma_L\!-\!\sigma_M)(\sigma_M\!-\!\sigma_R)+e^{2ik\sigma_L(c/\sigma_R+
b/\sigma_M)}(\sigma_L\!+\!\sigma_M)(\sigma_M\!-\!\sigma_R) \right.\\
&~~+\left.e^{2ick\sigma_L/\sigma_R}(\sigma_L-\sigma_M)(\sigma_M+\sigma_R)+e^{2ibk\sigma_L(1/\sigma_R+1/\sigma_M)}(\sigma_L+\sigma_M)(\sigma_M+\sigma_R)   \right)  \hat{u}_0^L(k) \ud k\\
&+\int_{\partial D_0^-}\!\!\!\!\!\!\!\!  \frac{e^{ik(x+2a)-(\sigma_Lk)^2t}}{2\Delta_L(k)} \!\left(\! e^{2ibk\sigma_L/\sigma_R}(\sigma_L\!-\!\sigma_M)(\sigma_M\!-\!\sigma_R)\!+\!e^{2ik\sigma_L(c/\sigma_R+
b/\sigma_M)}(\sigma_L\!+\!\sigma_M)(\sigma_M\!-\!\sigma_R)+\!\right.\\
&~~+\left.e^{2ick\sigma_L/\sigma_R}(\sigma_L-\sigma_M)(\sigma_M+\sigma_R)+e^{2ibk\sigma_L(1/\sigma_R+1/\sigma_M)}(\sigma_L+\sigma_M)(\sigma_M+\sigma_R)   \right)  \hat{u}_0^L(-k) \ud k\\
&-\int_{\partial D_0^-} \!\!\!\!\!\!\!\! \frac{e^{ik(x+2a+b\sigma_L/\sigma_M)-(\sigma_Lk)^2t}\sigma_L}{\Delta_L(k)}(e^{2ibk\sigma_L/\sigma_R}(\sigma_M\!
-\!\sigma_R)\!+\!e^{2ick\sigma_L/\sigma_R}(\sigma_M\!+\!\sigma_R))  \hat{u}_0^M(k\sigma_L/\sigma_M) \!\ud k\\
&-\int_{\partial D_0^-} \!\!\!\!\!\!\!\! \frac{ e^{ik(x+2a+b\sigma_L/\sigma_M)\!-\!(\sigma_Lk)^2t}\sigma_L}{\Delta_L(k)}(e^{2ick\sigma_L/\sigma_R}(\sigma_M\!-\!
\sigma_R)\!+\!e^{2ibk\sigma_L/\sigma_R}(\sigma_M\!+\!\sigma_R)) \hat{u}_0^M(-k\sigma_L/\sigma_M)\!\ud k \\
&+\int_{\partial D_0^-} \frac{2\sigma_L\sigma_M}{\Delta_L(k)} e^{ik(x+2a+2c\sigma_L/\sigma_R+b\sigma_L/\sigma_R+b\sigma_L/\sigma_M)-(\sigma_Lk)^2t} \hat{u}_0^R(k\sigma_L/\sigma_R)\ud k \\
&+\int_{\partial D_0^-} \frac{2\sigma_L\sigma_M}{\Delta_L(k)}e^{ik(x+2a+b\sigma_L/\sigma_R+b\sigma_L/\sigma_M)-(\sigma_Lk)^2t}  \hat{u}_0^R(-k\sigma_L/\sigma_R)\ud k \\
&+\int_{\partial D_0^+}\!\!\!\!\!\!\!\! \frac{e^{ik(x+2a)\!-\!(\sigma_Lk)^2t}}{2\Delta_L(k)}\!\left(\!e^{2ik\sigma_L(c/\sigma_R\!+\!b/\sigma_M)}
(\sigma_L\!-\!\sigma_M)(\sigma_M\!-\!\sigma_R)\!+\!e^{2ibk\sigma_L/\sigma_R}(\sigma_L\!+\!
\sigma_M)(\sigma_M\!-\!\sigma_R)\!+\!\right.\\
&~~+\left.e^{2ibk\sigma_L(1/\sigma_R+1/\sigma_M)}(\sigma_L-\sigma_M)(\sigma_M+\sigma_R)+e^{2ick\sigma_L/\sigma_R}(\sigma_L+\sigma_M)(\sigma_M+\sigma_R)    \right)  \hat{u}_0^L(k)\ud k \\
&+\int_{\partial D_0^+} \!\!\!\!\!\!\!\! \frac{e^{ik(x+2a)\!-\!(\sigma_Lk)^2t}}{2\Delta_L(k)} \left(e^{2ibk\sigma_L/\sigma_R}(\sigma_L\!-\!\sigma_M)(\sigma_M\!-\!\sigma_R)\!+\!e^{2ik\sigma_L(c/\sigma_R\!+\!
b/\sigma_M)}(\sigma_L\!+\!\sigma_M)(\sigma_M\!-\!\sigma_R) \right.\\
&~~+\left.e^{2ick\sigma_L/\sigma_R}(\sigma_L-\sigma_M)(\sigma_M+\sigma_R)+e^{2ibk\sigma_L(1/\sigma_R+1/\sigma_M)}(\sigma_L+\sigma_M)(\sigma_M+\sigma_R)       \right) \hat{u}_0^L(-k)\ud k \\
&-\int_{\partial D_0^+} \!\!\!\!\!\!\!\!\frac{e^{ik(x+2a+b\sigma_L/\sigma_M)\!-\!(\sigma_Lk)^2t}\sigma_L}{\Delta_L(k)}\!\left(\! e^{2ibk\sigma_L/\sigma_R}(\sigma_M\!-\!\sigma_R)\!+\!e^{2ick\sigma_L/\sigma_R}(\sigma_M\!+\!\sigma_R) \!\right)\!  \hat{u}_0^M(k\sigma_L/\sigma_M)\! \ud k\\
&-\int_{\partial D_0^+} \!\!\!\!\!\!\!\!\frac{e^{ik(x+2a+b\sigma_L/\sigma_M)\!-\!(\sigma_Lk)^2t}\sigma_L}{\Delta_L(k)}\!\left(\! e^{2ick\sigma_L/\sigma_R}(\sigma_M\!-\!\sigma_R)\!+\!e^{2ibk\sigma_L/\sigma_R}(\sigma_M\!+\!\sigma_R) \!\right)\!  \hat{u}_0^M(-k\sigma_L/\sigma_M)\! \ud k\\
&+\int_{\partial D_0^+}  \frac{2\sigma_L\sigma_M}{\Delta_L(k)}e^{ik(x+2a+2c\sigma_L/\sigma_R+b\sigma_L(1/\sigma_R+1/\sigma_M))-(\sigma_Lk)^2t} \hat{u}_0^R(k\sigma_L/\sigma_R)\ud k \\
&+\int_{\partial D_0^+}  \frac{2\sigma_L\sigma_M}{\Delta_L(k)}e^{ik(x+2a+b\sigma_L(1/\sigma_R+1/\sigma_M))-(\sigma_Lk)^2t} \hat{u}_0^R(-k\sigma_L/\sigma_R)\ud k ,
\end{align*}

\no for $-a<x<0$ with

\bea
\Delta_L(k)&=&\pi \left(e^{2ibk\sigma_L/\sigma_R}(\sigma_L-\sigma_M)(\sigma_M-\sigma_R)+
e^{2ik(c\sigma_L/\sigma_R+b\sigma_L/\sigma_M+a)}(\sigma_M-\sigma_L)(\sigma_M-\sigma_R) \right.\\
&&+e^{2ik\sigma_L(c/\sigma_R+b/\sigma_M)}(\sigma_L+\sigma_M)(\sigma_M-\sigma_R)+
e^{2ik(b\sigma_L/\sigma_R+a)}(\sigma_L+\sigma_M)(\sigma_R-\sigma_M)\\
&&+e^{2ick\sigma_L/\sigma_R}(\sigma_L-\sigma_M)(\sigma_M+\sigma_R)+
e^{2ik(a+b\sigma_L/\sigma_R+b\sigma_L/\sigma_M)}(\sigma_M-\sigma_L)(\sigma_M+\sigma_R)\\
&&\left.+e^{2ibk(\sigma_L/\sigma_R+\sigma_L/\sigma_M)}(\sigma_L+\sigma_M)(\sigma_M+
\sigma_R)-e^{2ik(c\sigma_L/\sigma_R+a)}(\sigma_L+\sigma_M)(\sigma_M+\sigma_R)\right).
\eea

\no Next,

\begin{align*}
&u^M(x,t)=\frac{1}{2\pi}\int_{-\infty}^\infty e^{i k x-(\sigma_M k)^2t}\hat{u}_0^M(k)\ud k\\
&\!+\!\int_{\partial D_0^-}   \frac{-e^{ik(x+b)-(\sigma_Mk)^2t}\sigma_M}{\Delta_M(k)} \left(e^{2ick\sigma_M/\sigma_R}(\sigma_M-\sigma_R)+e^{2ibk\sigma_M/\sigma_R}(\sigma_M+\sigma_R)   \right)  \hat{u}_0^L(k\sigma_M/\sigma_L) \ud k\\
&\!+\!\int_{\partial D_0^-} \!\!\!\!\!\! \frac{-e^{ik(x+b+2a\sigma_M/\sigma_L)-(\sigma_Mk)^2t}\sigma_M}{\Delta_M(k)} \!\left(\!e^{2ick\sigma_M/\sigma_R}(\sigma_M\!-\!\sigma_R)\!+\!e^{2ibk\sigma_M/\sigma_R}(\sigma_M\!+\!\sigma_R) \!\right)\!  \hat{u}_0^L(-k\sigma_M/\sigma_L)\! \ud k\\
&\!-\!\int_{\partial D_0^-}\!\!\!\!\! \frac{e^{ikx-(\sigma_Mk)^2t}}{2\Delta_M(k)}(\sigma_M\!-\!\sigma_L\!+\!
e^{2iak\sigma_M/\sigma_L}(\sigma_L\!+\!\sigma_M))(e^{2ibk\sigma_M/\sigma_R}(\sigma_M\!-\!\sigma_R)\!+\!
e^{2ick\sigma_M/\sigma_R}(\sigma_M\!+\!\sigma_R))  \hat{u}_0^M(k)\! \ud k\\
&\!-\!\int_{\partial D_0^-}\!\!\!\!\!\!  \frac{e^{ikx-(\sigma_Mk)^2t}}{2\Delta_M(k)}(\sigma_M\!-\!\sigma_L\!+\!
e^{2iak\sigma_M/\sigma_L}(\sigma_L\!+\!\sigma_M))(e^{2ick\sigma_M/\sigma_R}(\sigma_M\!-\!\sigma_R)\!+\!
e^{2ibk\sigma_M/\sigma_R}(\sigma_M\!+\!\sigma_R)) \hat{u}_0^M(-k)\!\ud k\\
&+\int_{\partial D_0^-} \frac{-e^{ik(x+\sigma_M/\sigma_R(b+2c))-(\sigma_Mk)^2t}\sigma_M}{\Delta_M(k)}(\sigma_M-\sigma_L+e^{2iak\sigma_M/\sigma_L}(\sigma_L+\sigma_M))    \hat{u}_0^R(k\sigma_M/\sigma_R)\ud k\\
&+\int_{\partial D_0^-}\frac{-e^{ik(x+b\sigma_M/\sigma_R)-(\sigma_Mk)^2t}\sigma_M}{\Delta_M(k)}(\sigma_M-\sigma_L+e^{2iak\sigma_M/\sigma_L}(\sigma_L+\sigma_M))   \hat{u}_0^R(-k\sigma_M/\sigma_R)\ud k\\
&+\int_{\partial D_0^+} \frac{-e^{ik(x+b)-(\sigma_Mk)^2t}\sigma_M}{\Delta_M(k)}\left(e^{2ick\sigma_M/\sigma_R}(\sigma_M-\sigma_R)+e^{2ibk\sigma_M/\sigma_R}(\sigma_M+\sigma_R)  \right)  \hat{u}_0^L(k\sigma_M/\sigma_L)\ud k\\
&\!-\!\int_{\partial D_0^+}\!\!\!\!\!\!  \frac{e^{ik(x+b+2a\sigma_M/\sigma_L)-(\sigma_Mk)^2t}\sigma_M}{\Delta_M(k)}
\!\left(\!e^{2ick\sigma_M/\sigma_R}(\sigma_M\!-\!\sigma_R)\!+\!e^{2ibk\sigma_M/\sigma_R}(\sigma_M\!+\!\sigma_R)  \!\right)\! \hat{u}_0^L(-k\sigma_M/\sigma_L)\!\ud k\\
&\!-\!\int_{\partial D_0^+}\!\!\!\!\!\!\!\! \frac{e^{ik(x+2b)-(\sigma_Mk)^2t}}{2\Delta_M(k)}
\!\left(\!\sigma_M\!+\!\sigma_L\!+\!e^{2iak\sigma_M/\sigma_L}(\sigma_M\!-\!\sigma_L)\!\right)\!
(e^{2ick\sigma_M/\sigma_R}(\sigma_M\!-\!\sigma_R)\!+\!e^{2ibk\sigma_M/\sigma_R}(\sigma_M\!+\!\sigma_R))  \hat{u}_0^M(k) \!\ud k\\
&\!-\!\int_{\partial D_0^+}\!\!\!\!\!\! \frac{e^{ikx\!-\!(\sigma_Mk)^2t}}{2\Delta_M(k)}\!\left(\!\sigma_M\!-\!\sigma_L\!+\!e^{2iak\sigma_M/\sigma_L}
(\sigma_M\!+\!\sigma_L)\!\right)\!(e^{2ick\sigma_M/\sigma_R}(\sigma_M\!-\!\sigma_R)\!+\!
e^{2ibk\sigma_M/\sigma_R}(\sigma_M\!+\!\sigma_R))  \hat{u}_0^M(-k)\! \ud k\\
&+\int_{\partial D_0^+}  \frac{-e^{ik(x+b\sigma_M/\sigma_R+2c\sigma_M/\sigma_R)-(\sigma_Mk)^2t}\sigma_M}{\Delta_M(k)}\left(\sigma_M-\sigma_L+e^{2iak\sigma_M/\sigma_L}(\sigma_L+\sigma_M)  \right)  \hat{u}_0^R(k\sigma_M/\sigma_R)\ud k+ \\
&+\int_{\partial D_0^+}  \frac{-e^{ik(x+b\sigma_M/\sigma_R)-(\sigma_Mk)^2t}\sigma_M}{\Delta_M(k)}\left(\sigma_M-\sigma_L+e^{2iak\sigma_M/\sigma_L}(\sigma_L+\sigma_M)  \right)    \hat{u}_0^R(-k\sigma_M/\sigma_R)\ud k ,
\end{align*}

\no for $0<x<b$, with

\bea
\Delta_M(k)&=&\pi \left(e^{ik(a\sigma_M/\sigma_L+b+c\sigma_M/\sigma_R)}(\sigma_L-\sigma_M)(\sigma_M-\sigma_R)+e^{2bik\sigma_M/\sigma_R}(\sigma_M-\sigma_L)(\sigma_M-\sigma_R) \right.\\
&&+e^{2ik\sigma_M(b/\sigma_R+a/\sigma_L)}(\sigma_L+\sigma_M)(\sigma_M-\sigma_R)+e^{2ik(c\sigma_M/\sigma_R+b)}(\sigma_L+\sigma_M)(\sigma_R-\sigma_M)\\
&&+e^{2ik(a\sigma_M/\sigma_L+b\sigma_M/\sigma_R+b)}(\sigma_L-\sigma_M)(\sigma_M+\sigma_R)+e^{2ick\sigma_M/\sigma_R}(\sigma_M-\sigma_L)(\sigma_M+\sigma_R)\\
&&\left.-e^{2ibk(\sigma_M/\sigma_R+1)}(\sigma_L+\sigma_M)(\sigma_M+\sigma_R)+e^{2ik\sigma_M(c/\sigma_R+a/\sigma_L)}(\sigma_L+\sigma_M)(\sigma_M+\sigma_R)\right),
\eea
and

\begin{align*}
&u^R(x,t)=\frac{1}{2\pi}\int_{-\infty}^\infty e^{i k x-(\sigma_Rk)^2t}\hat{u}_0^R(k)\ud k\\
&+\int_{\partial D_0^-}   \frac{-2e^{ik(x+b+b\sigma_R/\sigma_M)-(\sigma_Rk)^2t}}{\Delta_R(k)} \left(\sigma_M\sigma_R   \right)  \hat{u}_0^L(k\sigma_R/\sigma_L) \ud k\\
&+\int_{\partial D_0^-}  \frac{-2e^{ik(x+b+b\sigma_R/\sigma_M+2a\sigma_R/\sigma_L)-(\sigma_Rk)^2t}}{\Delta_R(k)} \left(\sigma_M\sigma_R      \right)  \hat{u}_0^L(-k\sigma_R/\sigma_L) \ud k\\
&+\int_{\partial D_0^-} \frac{e^{ik(x+b+b\sigma_R/\sigma_M)-(\sigma_Rk)^2t}}{\Delta_R(k)}(\sigma_L+\sigma_M+e^{2iak\sigma_R/\sigma_L}(\sigma_M-\sigma_L ))  \hat{u}_0^M(k\sigma_R/\sigma_M) \ud k\\
&+\int_{\partial D_0^-}  \frac{e^{ik(x+b)-(\sigma_Rk)^2t}}{\Delta_R(k)}(\sigma_M-\sigma_L+e^{2iak\sigma_R/\sigma_L}(\sigma_M+\sigma_L )) \hat{u}_0^M(-k\sigma_R/\sigma_M)\ud k \\
&+\int_{\partial D_0^-} \frac{e^{ik(x+2b)-(\sigma_Rk)^2t}}{2\Delta_R(k)}\left((\sigma_R-\sigma_M)(\sigma_L-\sigma_M-e^{2iak\sigma_R/\sigma_L}(\sigma_L+\sigma_M)) \right.\\
&~~~~~~~~\left.-e^{2ibk\sigma_R/\sigma_M}(\sigma_M+\sigma_R)(\sigma_L+\sigma_M+e^{2iak\sigma_R/\sigma_L}(\sigma_M-\sigma_L))  \right)    \hat{u}_0^R(k)\ud k \\
&+\int_{\partial D_0^-} \frac{e^{ikx-(\sigma_Rk)^2t}}{2\Delta_R(k)}\left((\sigma_R+\sigma_M)(\sigma_L-\sigma_M-e^{2iak\sigma_R/\sigma_L}(\sigma_L+\sigma_M)) \right.\\
&~~~~~~~~\left.-e^{2ibk\sigma_R/\sigma_M}(\sigma_R-\sigma_M)(\sigma_L+\sigma_M+e^{2iak\sigma_R/\sigma_L}(\sigma_M-\sigma_L))  \right)   \hat{u}_0^R(-k)\ud k \\
&+\int_{\partial D_0^+} \frac{2e^{ik(x+b+b\sigma_R/\sigma_M)-(\sigma_Rk)^2t}}{\Delta_R(k)}\left(\sigma_M\sigma_R     \right)  \hat{u}_0^L(k\sigma_R/\sigma_L)\ud k \\
&+\int_{\partial D_0^+}  \frac{2e^{ik(x+b+b\sigma_R/\sigma_M+2a\sigma_R/\sigma_L)-(\sigma_Rk)^2t}}{\Delta_R(k)} \left(  \sigma_M\sigma_R    \right) \hat{u}_0^L(-k\sigma_R/\sigma_L)\ud k \\
&+\int_{\partial D_0^+} \frac{-e^{ik(x+b+2b\sigma_R/\sigma_M)-(\sigma_Rk)^2t}\sigma_R}{\Delta_R(k)}\left( \sigma_L+\sigma_M+e^{2iak\sigma_R/\sigma_L}(\sigma_M-\sigma_L)  \right)  \hat{u}_0^M(k\sigma_R/\sigma_M) \ud k\\
&+\int_{\partial D_0^+} \frac{e^{ik(x+b)-(\sigma_Rk)^2t}\sigma_R}{\Delta_R(k)}\left(\sigma_M-\sigma_L+e^{2iak\sigma_R/\sigma_L}(\sigma_M+\sigma_L)   \right)  \hat{u}_0^M(-k\sigma_R/\sigma_M) \ud k\\
&+\int_{\partial D_0^+}  \frac{e^{ik(x+2c)-(\sigma_Rk)^2t}}{2\Delta_R(k)}\left( e^{2ibk\sigma_R/\sigma_M}(\sigma_R-\sigma_M)(\sigma_L+\sigma_M+e^{2iak\sigma_R/\sigma_L}) \right.\\
&~~~~~~~~\left.+(\sigma_M+\sigma_R)(\sigma_M-\sigma_L+e^{2iak\sigma_R/\sigma_L}(\sigma_L+\sigma_M))  \right)  \hat{u}_0^R(k)\ud k \\
&+\int_{\partial D_0^+}  \frac{e^{ikx-(\sigma_Rk)^2t}}{2\Delta_R(k)}  \left( e^{2ibk\sigma_R/\sigma_M}(\sigma_R-\sigma_M)(\sigma_L+\sigma_M+e^{2iak\sigma_R/\sigma_L}) \right.\\
&~~~~~~~~\left.(\sigma_M+\sigma_R)(\sigma_M-\sigma_L+e^{2iak\sigma_R/\sigma_L}(\sigma_L+\sigma_M))  \right) \hat{u}_0^R(-k)\ud k ,
\end{align*}

\no for $b<x<c$ with

\begin{align*}
\Delta_R(k)&=\pi \left(e^{2ibk}(\sigma_L-\sigma_M)(\sigma_M-\sigma_R)+e^{2ik(a\sigma_R/\sigma_L+b\sigma_R/\sigma_M+c)}(\sigma_M-\sigma_L)(\sigma_M-\sigma_R) \right.\\
&+e^{2ik(b\sigma_R/\sigma_M+c)}(\sigma_L+\sigma_M)(\sigma_M-\sigma_R)+e^{2ik(a\sigma_R/\sigma_L+b)}(\sigma_L+\sigma_M)(\sigma_R-\sigma_M)\\
&+e^{2ick}(\sigma_L-\sigma_M)(\sigma_M+\sigma_R)+e^{2ik(a/\sigma_L+b+b\sigma_R/\sigma_M)}(\sigma_M-\sigma_L)(\sigma_M+\sigma_R)\\
&\left.+e^{2ibk(1+\sigma_R/\sigma_M)}(\sigma_L+\sigma_M)(\sigma_M+\sigma_R)-e^{2ik(a\sigma_R/\sigma_L+c)}(\sigma_L+\sigma_M)(\sigma_M+\sigma_R)\right).
\end{align*}
\end{prop}


\begin{thebibliography}{1}

\bibitem{AblowitzFokas}
M.~J. Ablowitz and A.~S. Fokas.
\newblock {\em Complex variables: Introduction and Applications}.
\newblock Cambridge Texts in Applied Mathematics. Cambridge University Press,
  Cambridge, second edition, 2003.

\bibitem{CarslawJaeger}
H.~S. Carslaw and J.~C. Jaeger.
\newblock {\em Conduction of Heat in Solids}.
\newblock Oxford University Press, New York, 2nd edition, 1959.

\bibitem{DeconinckTrogdonVasan}
B.~Deconinck, T.~Trogdon, and V.~Vasan.
\newblock The method of {F}okas for solving linear partial differential
  equations.
\newblock {\em SIAM Rev.}, 56(1):159--186, 2014.

\bibitem{FlyerFokas}
N.~Flyer and A.~S. Fokas.
\newblock A hybrid analytical-numerical method for solving evolution partial
  differential equations. {I}. {T}he half-line.
\newblock {\em Proc. R. Soc. Lond. Ser. A Math. Phys. Eng. Sci.},
  464(2095):1823--1849, 2008.

\bibitem{FokasBook}
A.~S. Fokas.
\newblock {\em A unified approach to boundary value problems}, volume~78 of
  {\em CBMS-NSF Regional Conference Series in Applied Mathematics}.
\newblock Society for Industrial and Applied Mathematics (SIAM), Philadelphia,
  PA, 2008.

\bibitem{FokasPelloni4}
A.~S. Fokas and B.~Pelloni.
\newblock A transform method for linear evolution {PDE}s on a finite interval.
\newblock {\em IMA J. Appl. Math.}, 70(4):564--587, 2005.

\bibitem{GuentherLee}
R.~B. Guenther and J.~W. Lee.
\newblock {\em Partial differential equations of mathematical physics and
  integral equations}.
\newblock Dover Publications Inc., Mineola, NY, 1996.
\newblock Corrected reprint of the 1988 original.

\bibitem{HahnO}
D.~Hahn and M.~{\"O}zisik.
\newblock {\em Heat Conduction}.
\newblock John Wiley {\&} Sons, Inc., Hoboken, New Jersey, 3rd edition, 2012.

\end{thebibliography}
\end{document}